\documentclass[review]{elsarticle}

\usepackage{lineno,hyperref}
\usepackage{amsmath}
\usepackage{cleveref}
\usepackage{tikz}
\usepackage{subfigure}
\usepackage{booktabs} %for tables usage of midrule etc.
\hypersetup{hidelinks}
\journal{Computers \& Fluids}

\usepackage{numcompress}

\begin{document}

\begin{frontmatter}

\title{Evaluation of physics constrained data-driven methods for turbulence model uncertainty quantification}

\author{Marcel Matha \corref{mycorrespondingauthor}}
\ead{marcel.matha@dlr.de}
\author{Karsten Kucharczyk and Christian Morsbach}
\address{German Aerospace Center (DLR), Linder Höhe, 51147 Cologne, Germany}
\cortext[mycorrespondingauthor]{Corresponding author}

\begin{abstract}
In order to achieve a virtual certification process and robust designs for turbomachinery, the uncertainty bounds for Computational Fluid Dynamics have to be known. The formulation of turbulence closure models implies a major source of the overall uncertainty of Reynolds-averaged Navier-Stokes simulations. We discuss the common practice of applying a physics constrained eigenspace perturbation of the Reynolds stress tensor in order to account for the model form uncertainty of turbulence models. Since the basic methodology often leads to overly generous uncertainty estimates, we extend a recent approach of adding a machine learning strategy.
The application of a data-driven method is motivated by striving for the detection of flow regions, which are prone to suffer from a lack of turbulence model prediction accuracy. In this way any user input related to choosing the degree of uncertainty is supposed to become obsolete. This work especially investigates an approach, which tries to determine an a priori estimation of prediction confidence, when there is no accurate data available to judge the prediction.
The flow around the NACA 4412 airfoil at near-stall conditions demonstrates the successful application of the data-driven eigenspace perturbation framework. Furthermore, we especially highlight the objectives and limitations of the underlying methodology.

\end{abstract}


\end{frontmatter}

%\linenumbers

\section{Introduction}
\label{introduction}
 \noindent In previous times, engineering design applications tried to account for a variety of uncertainties in Computational Fluid Dynamics (CFD) simulations by applying factors of safety, margins of safety, levels of redundancy, etc. Such heuristic methods need adjustments and recalibration for each new configuration. They are also highly empirical especially when applied to innovative designs or new flow configurations.
  Nowadays methodologies such as robust design or reliability based design have the possibility to preempt such methods. 
  In recent years, the interest in uncertainty quantification (UQ), leading to more reliable simulation based designs, has grown significantly \cite{Karniadakis, Oberkampf, EmoryThesis, XIAO20191}.\\
As a compromise between computational time and accuracy, Reynolds-averaged Navier-Stokes (RANS) based turbulence modeling is still the prevailing tool in industrial design of turbomachinery, since the replacement of RANS by scale-resolving simulations, e.g. Direct Numerical Simulations (DNS) or Large Eddy Simulations (LES) cannot be expected for design optimization simulations in the near future.
 The derivation of the RANS equations reveals an unclosed term, called the Reynolds stress tensor. This tensor has to be approximated in CFD simulations by applying turbulence models. 
  The prediction quality of the simulation is highly dependent on the accuracy of the turbulence models.
  However, many RANS-based models suffer from the inability to replicate fundamental turbulent processes. Throughout this paper, we consider linear eddy viscosity models (LEVM), which are widely used for complex engineering flows, by referring to RANS turbulence models.
  Due to simplifying assumptions used in model formulation, the turbulence model is one of the main limitations in striving for the next generation of reliable, efficient and environmentally friendly designs.
  These simplifications are the result of data observation, physical intuition, engineering and computational pragmatism, leading to a significant degree of epistemic uncertainty. 
  Accounting for the uncertainties, which arise due to the structural form of the turbulence model in RANS simulation is known to be the "greatest challenge" in CFD \cite{Zang}. Nevertheless, these epistemic uncertainties could, in principle, be reduced, by increasing knowledge about turbulent processes, resulting in developing advanced models. This is contrary to aleatory uncertainties, e.g. manufacturing tolerances or operating conditions, which cannot be reduced and are not considered in the current work.
  Different approaches try to account for the uncertainty of the turbulence model at different modelling levels \cite{Duraisamy}. Generally, one distinguishes between parametric and non-parametric approaches. While the parametric uncertainties arise from the chosen closure coefficients and their calibration process, non-parametric methodologies directly investigate the uncertainties on modeled terms in the transport equations of the turbulence model. It is expected, that the possible solution space, with respect to the uncertainty of the turbulence model, is larger for non-parametric approaches \cite{XIAO20191}.\\
  Iaccarino and co-workers proposed an eigenspace perturbation framework, which is based on the inability of common LEVM to deal with Reynolds stress tensor anisotropy \cite{Emory, Iaccarino}. This methodology belongs to the non-parametric approaches, since it tries to account for the uncertainty due to the closure model form itself. The physical rationale of the eigenspace perturbation framework is discussed in depth by Mishra and Iaccarino \cite{Mishra2019}.
  It enables a designer to optimize components towards an optimum design with less sensitivity to uncertainty. Successful application of design under uncertainty using the perturbation framework was already presented by Mishra et al. \cite{Mishra2020Design}.
  Moreover, this methodology was applied in multiple engineering applications such as aerospace design (aircraft nozzle \cite{Iaccarino}, turbomachinery \cite{EmoryTurbo, Razaaly}, entire aircraft \cite{Mukhopadhaya}), civil structural design \cite{Lamberti} and even wind farm design \cite{EidiDataFree, Hornshoj}.\\
  The increasing availability of high-fidelity simulations (such as LES and DNS) in combination with the emergence of machine learning strategies guided the path towards data-driven approaches also for the RANS turbulence modelling community \cite{Duraisamy}. 
  Heyse et al. \cite{Heyse} enhanced the uncertainty estimation based on the eigenspace perturbation approach by adding a data-driven method leading to less conservative uncertainty estimates. The machine learning strategy should identify flow regions, which are prone to show reduced turbulence model prediction accuracy. From our point of view, their investigations suffered under limited availability of data, most notably with respect to judging an appropriate application of a machine learning model. We consolidate the usage of a random forest regression model by investigating the credibility explicitly in the present work.\\
  %In our previous work, we already demonstrated the possibility to use the proposed data-driven eigenvalue perturbation approach by Heyse et al. \cite{Heyse} and validated our implementation in DLR's CFD solver suite TRACE for turbulent channel flow \cite{MathaNeurIPS}.
  In the current work, we present the functionality of the eigenspace perturbation methodology, its data-driven extension and its implementation within DLR's CFD solver suite TRACE in detail.
  TRACE is being developed by the Institute of Propulsion Technology with focus on turbomachinery flows and offers a parallelized, multi-block CFD solver for the compressible RANS equations \cite{morsbach_RSM}. 
  In order to obtain a trustworthy quantification of uncertainties for future design application in turbomachinery flows, we investigate our implementation of the data-driven Reynolds stress tensor perturbation framework for flow configurations featuring adverse pressure gradient, flow separation and reattachment with TRACE.
  This procedure is reasonable, as the prediction accuracy of RANS turbulence models is significantly reduced in the presence of these complex flow phenomena.
  In order to further advance the data-driven prediction capabilities, additional data sets obtained by scale-resolving simulations of relevant test cases are used to train and validate a machine learning model. These test cases include complex flow physics such as adverse pressure gradient, separation and reattachment.
  In this work, we verify the application of a trained machine learning model in detail. A methodology to quantify an a priori estimate of prediction confidence is particularly studied as well.
  Finally, the data-driven perturbation approach to estimate the epistemic uncertainty of turbulence models is applied for the flow around the NACA 4412 airfoil at near-stall conditions featuring a separation zone on the suction surface. By analyzing and comparing the uncertainty estimates for certain quantities of interest (QoI) with respect to the usage of the data-free and data-driven strategy, we analyze the initial intention of the entire Reynolds stress tensor perturbation framework and its capabilities consequently. From our point of view, this kind of subsumption was missing in literature oftentimes.

\section{Details of the eigenspace perturbation framework}
\label{sec_eigenspaceFramework}

\subsection{Motivation and goal}
\label{motivationPerturbation}

\noindent RANS turbulence models are utilized in order to approximate the Reynolds stress tensor $\tau_{ij} = \overline{u_i' u_j'}$ in terms of mean flow quantities, e.g. $\overline u_i = U_i = u_i - ~ u_i'$. As already described, the formulation of turbulence models brings along certain assumptions. Even state-of-the-art LEVM rely for example on the eddy viscosity hypothesis, also known as the Boussinesq assumption, and the gradient diffusion hypothesis. This leads to the inability to account for rotational effects, secondary flow, swirl and streamline curvature \cite{Speziale, Mompean, CRAFT1996108}. Besides, limited success in cases with separation and reattachment have been also reported \cite{Lien}. The motivation for injecting perturbations to the eigenspace of the Reynolds stress tensor is the inability of LEVM to account correctly for the anisotropy of Reynolds stresses. This is due to the Boussinesq assumption, approximating the turbulent stresses in similar manner to the molecular viscous stresses. The Boussinesq approximations reads
\begin{linenomath}
\begin{equation}
\label{eq:boussinesq}
    \tau_{ij} = -2 \mu_t \left(S_{ij} - \frac{1}{3}\frac{\partial u_k}{\partial x_k}\delta_{ij}\right) + \frac{2}{3} k \delta_{ij} \ \text{,}
\end{equation}
\end{linenomath}
where the turbulent kinetic energy is defined as $k = \frac{1}{2} \tau_{kk}$ and summation over recurring indices within a product is implied.
The strain-rate tensor is denoted as $S_{ij}$ and the eddy viscosity, derived from the transport equations of the turbulence model, is represented by $\mu_t$.\\
Based on the epistemic uncertainty, which is introduced into turbulence models by the choice of the actual closure model \cite{Duraisamy}, the perturbation approach tries to derive and quantify the effects on QoI, e.g. the pressure field, by modifying the anisotropy of turbulence within physical limitations. 
The implemented framework for UQ of turbulence models seeks to sample from solutions, resulting from a modified underlying structure of the turbulence model, while aiming for extreme states of the Reynolds stress tensor \cite{Mishra2019}. In this manner a CFD practitioner may get the chance to estimate the sensitivity of some QoI regarding the uncertainty in predicting the Reynolds stress anisotropy.
In the following section we explain how to obtain a perturbed state of the Reynolds stress tensor and how to apply machine learning in order to get a better-informed, less conservative uncertainty prediction.

\subsection{Data-free approach}
\label{sec_dataFree}

\begin{figure}[h!]
\begin{center}
\mbox{\subfigure[Unperturbed and perturbed state]{\includegraphics[width=0.49\textwidth, trim=0.cm 0.18cm 0.cm 0.cm, clip=True]{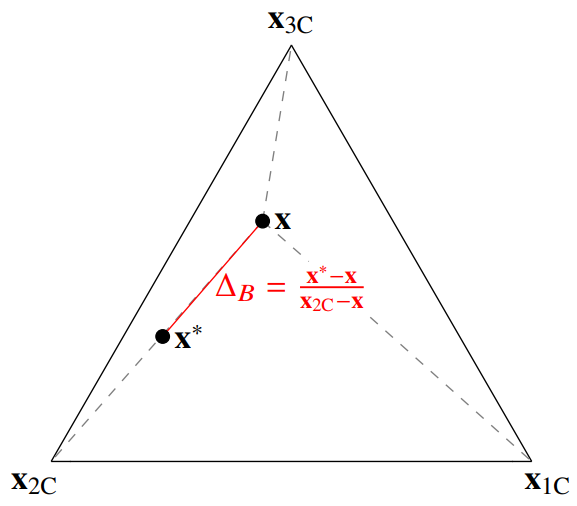}\label{baryCentric1}}}
\mbox{\subfigure[Definition of perturbation strength]{\includegraphics[width=0.49\textwidth]{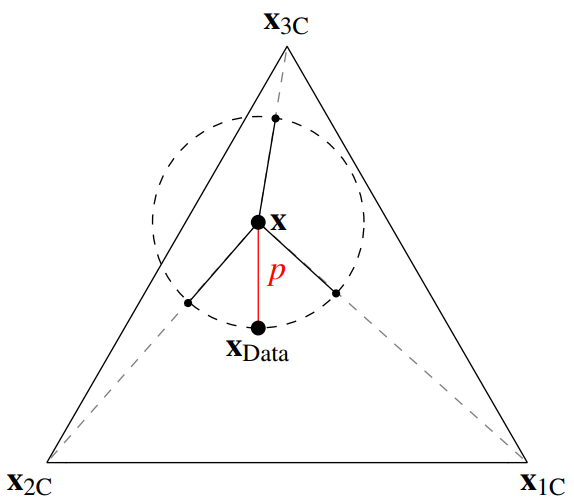}\label{baryCentric2}}}
\vspace{-0.5\baselineskip}
\caption{Schematic representation of the eigenvalue perturbation approach}
\label{Bild:barycentricTriangle}
\end{center}
\end{figure}

\noindent The symmetric Reynolds stress tensor can be expressed by applying an eigenspace decomposition as
\begin{linenomath}
\begin{equation}
	\label{spectralDecompositionR}
		\tau_{ij} = k \left(a_{ij} + \frac{2}{3}\delta_{ij}\right) = k \left(v_{in} \Lambda_{nl} v_{jl} + \frac{2}{3}\delta_{ij}\right) \ \text{.}
\end{equation} 
\end{linenomath}
Equation \eqref{spectralDecompositionR} includes the split into the anisotropy tensor $a_{ij}$ and the isotopic part of $\tau_{ij}$.
The eigenspace decomposition provides the eigenvector matrix $v$ and the diagonal eigenvalue matrix $\Lambda$, where the eigenvalues represent the shape and the eigenvectors imply the orientation of the tensor.
Emory et al. \cite{Emory} propose a strategy to perturb the eigenvalues and eigenvectors in Equation \eqref{spectralDecompositionR}, resulting in a perturbed state of the Reynolds stress tensor
\begin{linenomath}
\begin{equation}
	\label{spectralDecompositionR*}
		\tau_{ij}^* = k \left(v_{in}^* \Lambda_{nl}^* v_{jl}^* + \frac{2}{3}\delta_{ij}\right) \ \text{.}
\end{equation}
\end{linenomath}
The eigenvalue perturbation (determining $\Lambda^*$) makes use of the fact, that every physical, realizable state of the Reynolds stress tensor can be mapped onto barycentric coordinates
\begin{linenomath}
\begin{equation}
\label{barycentricMapping}
	\mathbf{x} = \mathbf{x}_{1C}\frac{1}{2}\left(\lambda_1-\lambda_2\right)+ \mathbf{x}_{2}\left(\lambda_2-\lambda_3\right)+ \mathbf{x}_{3C} \left(\frac{3}{2}  \lambda_3+1\right) \quad \text{with} \quad \lambda_1\geq\lambda_2 \geq\lambda_3 \ \text{,}
\end{equation}
\end{linenomath}
which is essentially a linear transformation according to $\mathbf{x} = \mathbf{B} \boldsymbol{\lambda}$ (whereby three eigenvalues $\lambda_i$ are represented by the vector $\boldsymbol{\lambda}$ and $\mathbf{B}$ depends on the choice of coordinates $\mathbf{x}_{\mathrm{1C}}, \mathbf{x}_{\mathrm{2C}}$ and $\mathbf{x}_{\mathrm{3C}}$) \cite{Banerjee2007}. \Cref{baryCentric1} shows the three limiting states of the Reynolds stress tensor, represented by the corners of the triangle ($\mathbf{x}_{\mathrm{1C}}, \mathbf{x}_{\mathrm{2C}}, \mathbf{x}_{\mathrm{3C}}$) standing for the one-, two- and three-component (isotropic) turbulent state ($\mathrm{1C}$, $\mathrm{2C}$ and $\mathrm{3C}$).
Thus, Iaccarino and co-workers \cite{Emory, Iaccarino} defined the eigenvalue perturbation as a shift in barycentric coordinates towards each of the limiting states to location $\mathbf{x}^*$, according to 
\begin{linenomath}
\begin{equation}
	\label{perturbationMagnitude}
		\mathbf{x}^* = \mathbf{x} + \Delta_B \left(\mathbf{x}_{(t)} -\mathbf{x}\right) \ \text{.}
\end{equation}
\end{linenomath}
The relative distance $\Delta_B \in [0, 1]$ controls the magnitude of eigenvalue perturbation towards the corner state $\mathbf{x}_{(t)} \in \{\mathbf{x}_{\mathrm{1C}}, \mathbf{x}_{\mathrm{2C}}, \mathbf{x}_{\mathrm{3C}}\}$.
The perturbed eigenvalues $\lambda_{i}^*$ can be remapped by
\begin{linenomath}
\begin{equation}
	\label{perturbedEigenvalues}
		\boldsymbol{\lambda}^* = \mathbf{B}^{-1} \mathbf{x}^* \ \text{.}
\end{equation}
\end{linenomath}

\noindent The creation of the perturbed eigenvector matrix $v^*$ is purely motivated by manipulating the turbulent production term $P_k=-\tau_{ij}\frac{\partial u_i}{\partial x_j}$. Changing the alignment of the eigenvectors of the Reynolds stress tensor and the strain rate tensor $S_{ij}$ limits the production term to a maximum and minimum value \cite{Iaccarino}. Maximum turbulent production is obtained by not changing the eigenvectors of the Reynolds stress tensor, meaning that they are identical to the eigenvectors of the strain rate tensor $v_{k_{S_{ij}}}$ due to the Boussinesq assumption in Equation \eqref{eq:boussinesq}. Commuting the first and the last eigenvector of the Reynolds stress tensor leads to minimum turbulent production:
\begin{linenomath}
\begin{align}
    \begin{split}
        \label{eq:minMaxProduction}
        v_\textrm{max} &= 
        \begin{pmatrix}
         v_{1_{S_{ij}}} & v_{2_{S_{ij}}} & v_{3_{S_{ij}}} 
        \end{pmatrix} \rightarrow P_{k_\textrm{max}} \\
        v_\textrm{min} &= 
        \begin{pmatrix}
        v_{3_{S_{ij}}} & v_{2_{S_{ij}}} & v_{1_{S_{ij}}} 
        \end{pmatrix} \rightarrow P_{k_\textrm{min}} \ \text{.}
    \end{split}
\end{align}
\end{linenomath}
When combining the eigenvalue and eigenvector perturbation, not only the shape of the Reynolds stress ellipsoid is modified but also the relative alignment with the principle axes of the mean rate of the strain rate tensor is changed (orientation).
In should be noted, that targeting the $\mathrm{3C}$ turbulent state with $\Delta_B =1$ results in identical eigenvalues and consequently the eigenvector matrix cancels out with its inverse. That is the reason, why there will be no distinction between minimized and maximized turbulent production.\\
To sum up, the data-free perturbation framework promises to only need five distinct simulations $\in \{(\mathrm{1C}, P_{k_\textrm{max}})$, $(\mathrm{1C}, P_{k_\textrm{min}})$, $(\mathrm{2C}, P_{k_\textrm{max}})$, $(\mathrm{2C}, P_{k_\textrm{min}})$ and $\mathrm{3C}\}$, in order to get the entire information with reference to the epistemic uncertainty of the underlying turbulence model, if $\Delta_B = 1$ is chosen.

\subsection{Data-driven approach}
\label{sec_dataDriven}

\noindent The eigenspace perturbation approach is a purely physics-based methodology, aiming for understandable uncertainty bounds for the turbulence modelling community. The data-free procedure applies a uniform perturbation to the entire flow domain.
But the perturbation amplitude is a reflection of the inability of the turbulence model to reflect the underlying physics with high fidelity. This discrepancy between the turbulence model's dynamics and those of the turbulence physics are not uniform, but differ between different turbulent flows and even across different regions of the flow domain in the same flow. Thus, enabling variation in the magnitude of the perturbations is a better reflection of the actual model form uncertainty. As an additional advantage, if executed correctly, such a varying perturbation approach would enable more precise and less conservative uncertainty bounds on the QoIs.
Moreover, a user has to choose the degree of uncertainty by selecting $\Delta_B$ before each investigation, which is another major drawbacks of the proposed method.
This might be especially unfavorable in the design phase of turbomachinery components, when even a CFD practitioner experienced in turbulence modeling does not know how to set the degree of uncertainty.\\
Consequently, the user defined bounds on the eigenspace perturbation procedure need to be replaced. 
Data-driven modeling approaches can be very beneficial for such surrogate modeling. Such machine learning surrogate models have found wide application in turbulence modeling \cite{Duraisamy, ihme2022combustion, brunton2020machine}. As the amount of high-fidelity simulations (such as LES and DNS) increases, this data can be used to estimate the perturbation magnitudes. 
Heyse et al. propose a strategy to obtain a locally varying perturbation strength by using a random forest model \cite{Heyse}. Physical flow features are extracted to train a machine learning model in order to predict the local perturbation strength 
\begin{linenomath}
\begin{equation}
    \label{eq:perturbationMagnitudeP}
    p = |\mathbf{x}_{\mathrm{Data}} - \mathbf{x}_{\mathrm{RANS}}| = |\mathbf{x}^*_{\mathrm{RANS}} - \mathbf{x}_{\mathrm{RANS}}| \ \text{,}
\end{equation}
\end{linenomath}
as illustrated in \Cref{baryCentric2}.
Forward propagating CFD simulations follow training the model, where the predicted perturbation strength is used to modify the Reynolds stress towards the same three limiting states as in the data-free approach. \\
In this work, we go several steps beyond the initial proposed strategy. 
First of all, Heyse et al. \cite{Heyse} applied pure eigenvalue perturbation of the anisotropy tensor. We combine the data-driven eigenvalue perturbation with the data-free manipulation of the eigenvectors of the Reynolds stress tensor, as already described above. From our point of view, this procedure is inevitable, since the assumption that the Reynolds stress tensor and the mean rate of strain tensor share their eigenvectors (see Equation \eqref{eq:boussinesq}), is known to be invalid in flow situations, where streamline curvature, rotational effects, flow separation or reattachment, etc play a role \cite{Launder}.
We decided against utilizing machine learning to adjust the perturbations for the eigenvectors of the Reynolds stress tensor and the turbulent kinetic energy directly, as this would lead to a data-augmented, corrected turbulence model instead of obtaining uncertainty estimates. Since no justifiable bounds for the turbulent kinetic energy budget exist, the present methodology perturbs it indirectly by applying the described eigenvector manipulation.   
Furthermore, we evaluate the adequate usage of a machine learning model for the desired application by extending the training data set and performing certain verification checks. The latter also includes the question, how to build trust in such a trained machine learning model, when there is no accurate data available anymore to estimate the prediction error.
And finally, we assess the limitations and capabilities of this method.

\subsubsection{Choice of machine learning model}
\noindent The general concept of machine learning is to approximate the relationship between input quantities (features) and output quantities (targets) to make prediction under similar conditions.
There are multiple approaches to approximate these relationships.
For the sake of interpretability and usability, decision trees are chosen to be the machine learning model in the present work. 
Decision trees (also called regression trees for solving regression problems) learn binary rules (if/else decision rules) to predict target values based on given features \cite{Breiman}.
Decision trees are prone to overfitting, which means that the model is not able to generalize. A machine learning model is able to generalize, if it performs adequate predictions based on a feature space, that is different than the feature space of the training data. 
Machine learning models, which are prone to overfitting reveal a high variance.
The potential accuracy of a machine learning model is also dependent on its bias, which is characterized by the difference between the averages of the predictions and the true values. An inflexible model is not capable to fit the total number of data sufficiently, which is determined as a high bias of the model.
The fact, that an increasing flexibility (lower bias) comes along with worse generalization (high variance), is known as the bias-variance trade-off.
This trade-off describes the aim to choose a machine learning model, that has low variance and low bias simultaneously \cite{Geman1992, Gilles2014}.
Random forests are based on a number of uncorrelated regression trees and offer the possibility to handle the bias-variance trade-off, while enabling powerful predictions \cite{Gilles2014}. For this reason, we have chosen to use this ensemble learning technique.
Instead of just averaging the prediction of individual regression trees, a random forest makes use of two essential key concepts:
\begin{itemize}
    \item bootstrapping: Random sampling (with replacement) of the training data for each individual tree, i.e. each tree is trained on a different data set with equal size.
    \item feature bagging: Random subsampling of features at each decision point (also known as split) for each tree, i.e. every tree uses a different feature space at each binary decision.
\end{itemize}
In combination with bootstrap aggregation (bagging), which implies averaging the prediction of a number of bootstrapped regression trees, the variance is reduced and overfitting is avoided \cite{Breiman2004BaggingP}.
In this work, the python library \textit{scikit-learn} \cite{scikit-learn} is used to train the random forests and evaluate their predictions.

\subsubsection{Choice of flow features}
\noindent The selection of input features, which are relevant for predicting more accurate perturbation magnitude $p$, is critical for turbulence modelling purposes.
It has to be ensured, that the chosen features represents physical significance with respect to desired target quantity.
Wang et al. \cite{Wang} identified four raw quantities 
\begin{linenomath}
\begin{equation}
    Q = \left(\mathbf{S}, \mathbf{\Omega}, \nabla p, \nabla k\right)
\end{equation}
\end{linenomath}
to be a reasonable choice as input data for conducting machine learning based on LEVM.
The two raw input tensors $\mathbf{S}$, $\mathbf{\Omega}$ represent strain rate and rotation rate, while $\nabla p$ and $\nabla k$ are the gradients of pressure and turbulent kinetic energy.
In our work, we agree on the usage of $Q$, and make use of the normalization scheme, derived by Ling et al. \cite{Ling}.
A normalization by a factor $\beta$ and the absolute value of each element $\alpha$ of $Q$ according to
\begin{linenomath}
\begin{equation}
\label{eq:normalizationFeatures}
    \hat{\alpha} = \frac{\alpha}{|\alpha|+|\beta|} \ \text{,}
\end{equation}
\end{linenomath}
lead to the determination of non-dimensional raw flow features, which are presented in \Cref{tab:rawFlowFeatures}.

\begin{table}[htb]
  \caption{Raw flow features for constructing the invariant basis}
  \label{tab:rawFlowFeatures}
  \centering
  \hspace*{-0.5cm}
  {\footnotesize
  \begin{tabular}{c c c c}
    \toprule
    Description&Normalized input $\hat{\alpha}$& raw input $\alpha$ & normalization factor $\beta$  \\
    \midrule
    Strain rate &$\mathbf{\hat{S}}$      &$\mathbf{S}$         &  $\omega$\\
    Rotation rate&$\mathbf{\hat{\Omega}}$ &$\mathbf{\Omega}$    & $||\mathbf{\Omega}||$\\
    Pressure gradient&$\widehat{\nabla p}$        &$\nabla p$           & $\rho ||\mathbf{U} \cdot \nabla \mathbf{U}||$\\
    Turbulent kinetic energy gradient&$\widehat{\nabla k}$        &$\nabla k$          & $\omega \sqrt{k}$\\
    \bottomrule
  \end{tabular}
  }
\end{table}

\noindent Since we are aiming for Galilean invariant features, which means, that they should stay the same in different inertial frames of reference, it is essential to embed invariance properties into a machine learning model (if the model should not learn these properties during training).
One method to do this is by formulating all inputs and output quantities of the model such that they are invariant. 
The velocity gradients (its symmetric part $\mathbf{\hat{S}}$ and antisymmetric part $\mathbf{\hat{\omega}}$) and the gradients of pressure and turbulent kinetic energy are Galilean invariant.
In order to determine the invariant feature basis of the raw flow features, Wang et al. \cite{Wang} make use of the Hilbert basis theorem. 
This theorem states, that a finite number of invariants belongs to each minimal integrity basis for a finite tensorial set \cite{Hilbert}.
A minimal integrity basis is the minimal set of invariants, that represent all
polynomial invariants associated with a tensorial set under transformation. The integrity basis constructed here is supposed to be rotational invariant \cite{Wu}.
In this manner the minimal integrity basis amount to 47 invariants, which are in the following used as input features for training and evaluating the random forest.
We add additional physical meaningful flow features to this exhaustive list of features based on domain knowledge and physical intuition. The additional raw input features, which are presented in \Cref{tab:physicalFeature}, can be computed by providing the turbulent kinetic energy $k$, the specific turbulent dissipation rate $\omega$, the molecular viscosity $\mu$, the eddy viscosity $\mu_t$, the distance to the nearest wall $d$, the local Mach number $Ma$, the mean velocity $U_i$ and its gradient tensor and the mean pressure $p$ and its gradient vector. 

\begin{table}[htb]
  \caption{Physical flow features}
  \label{tab:physicalFeature}
  \centering
  \hspace*{-0.9cm}
  {\footnotesize
  \begin{tabular}{c c c c}
    \toprule
    Numbering&Description & raw input $\alpha$ & normalization factor $\beta$  \\[1ex]
    \midrule
    $q_1$&Q-criterion      &$\frac{1}{2}\left(||\mathbf{\Omega}||^2 - ||\mathbf{S}||^2\right)$         &  $||\mathbf{S}||^2$\\[1ex]
    $q_2$&Turbulent kinetic energy & $k$& $\frac{1}{2}U_i U_i$\\[1ex]
    $q_3$&Wall-distance based Reynolds Number        &$\min\left(\frac{\sqrt{k}d}{50\nu},2\right)$           & -\\[1ex]
    $q_4$&Pressure gradient along streamline        &$U_k\frac{\partial p}{\partial x_k}$          & $\sqrt{\frac{\partial p}{\partial x_j}\frac{\partial p}{\partial x_j}U_iU_i}$\\[1ex]
    $q_5$&Turbulent time scale        &$\frac{1}{\omega}$          & $\frac{1}{||\mathbf{S}||}$\\[1ex]
    $q_6$&Production term        &$P_k$          & $k\omega$\\[1ex]
    $q_7$&Mach number        & $Ma$          & -\\[1ex]
    $q_8$&Eddy viscosity        & $\mu_t$          & $\mu$\\[1ex]
    $q_9$&Norm of Reynolds stresses        & $||\overline{u_i'u_j'}||$        & k\\
    \bottomrule
  \end{tabular}
  }
\end{table}
\noindent The normalization procedure is retained in accordance to Equation \eqref{eq:normalizationFeatures}. It should be noted, that the raw inputs and the normalization factors are Galilean invariant as well. Thus, a total number of 56 input features is used for training and evaluating the random forests.
Lastly, each feature is standardized by removing the mean and scaling to unit variance by applying a standard scaler preprocessing functionality of \textit{scikit-learn} \cite{scikit-learn}.\\
Although Wang et al. \cite{Wang_2018} reported significant improvements in prediction accuracy for using the invariant feature basis over a smaller number of features (e.g. physical motivated scalars) in a comparable study, we cannot confirm such observations in our work. We rather think, that due to the limited number of data (for training and testing purposes) a smaller number of input quantities performs excellently as well. As soon as the diversity of the data starts to increase (significantly different geometries featuring various flow situations) the need for a large set of features may occur. Being aware of the fact, that the present feature list may be needlessly large, we assure, that this exhaustive list of features does not involve any disadvantages in terms of accuracy. For this reason, we use the total amount of 56 input features in the present work.

\subsection{Integration of UQ computation in CFD solver suite TRACE}

\subsubsection{Implementation}
\label{sec:implementation}
\noindent The aim of running a CFD simulation, propagating a perturbed Reynolds stress tensor, is to obtain a sensitivity with respect to the solution. For smooth and time-efficient simulations, it is advisable to start the perturbation from a sufficiently converged baseline RANS simulation (baseline means standard unmodified turbulence model). Mishra et al. \cite{MishraSU2} apply a factor to march the solution based on the perturbed Reynolds stress tensor to a fully converged state.  In our implementation, we use a factor $f$ for the reconstruction of Reynolds stresses in order to be able to achieve fully converged perturbed solutions as well. We discuss the necessity and the effect of this factor in \Cref{sec:discussionRestrictions}.
The perturbation of the Reynolds stress tensor was implemented to the existing C code of TRACE and can be subdivided in several steps within each pseudo-time step of steady simulations:
\begin{enumerate}
    \item \label{item:step1} calculate Reynolds stresses based on Boussinesq approximation in Equation \eqref{eq:boussinesq}
    \item determine anisotropy tensor (see Equation \eqref{spectralDecompositionR})
    \item perturb anisotropy tensor within physical realizable limits by selecting $\Delta_B$ (see Equation \eqref{perturbationMagnitude}) and whether the turbulent production term should be minimized or maximized (see Equation \eqref{eq:minMaxProduction})
    \item reconstruct perturbed Reynolds stress tensor according to 
    \begin{linenomath}
    \begin{equation}
        \label{reconstructionReStress}
        \tau_{{ij}_f}^* = \tau_{ij} + f \left[k \left(a^* + \frac{2}{3}\delta_{ij}\right) - \tau_{ij}\right] \text{,}
    \end{equation}
    \end{linenomath}
    where $f \in [0, 1]$ is the introduced
    moderation factor, adjusting the total amount of newly perturbed anisotropy tensor to be considered
    \item update of the viscous fluxes using perturbed Reynolds stresses explicitly 
    \item update of the turbulent production term $P_k=-\tau_{ij}\frac{\partial u_i}{\partial x_j}$ using the perturbed Reynolds stresses explicitly
\end{enumerate}

\begin{figure}[htb]
\begin{center}
\includegraphics[width=\textwidth]{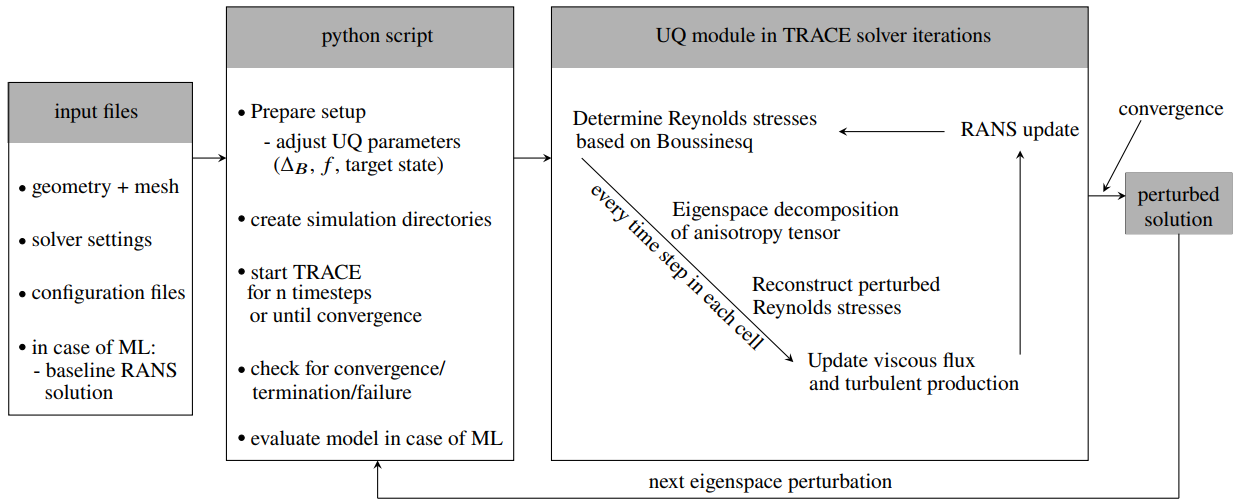}
\caption{Implementation of the UQ framework within CFD solver suite TRACE}
\label{Bild:TRACEuq}
\end{center}
\end{figure}

\noindent TRACE features a python interface, called pyTRACE \cite{TRACEUserGuide}, which can be used to conduct a full set of perturbed simulations and sample the results for some QoI. In case of applying a data-driven perturbation of the Reynolds stresses, the python script takes also charge of evaluating a previously in preprocessing trained machine learning model based on extracted mean flow quantities.
The high-level python script takes input parameters, containing information regarding the geometry, mesh resolution and additional solver settings. Furthermore, the set of intended perturbed simulations is set up, including selecting the turbulence limit state ($\mathbf{x}_{(t)} \in \{\mathbf{x}_{\mathrm{1C}}, \mathbf{x}_{\mathrm{2C}}, \mathbf{x}_{\mathrm{3C}}\}$), the relative distance $\Delta_B \in [0, 1]$, the alignment of the Reynolds stress eigenvectors with the strain rate tensor ($v_\textrm{min}$ or $v_\textrm{max}$) and the moderation factor $f \in [0, 1]$, as described earlier.
As illustrated in \Cref{Bild:TRACEuq} the integration of the UQ module in the TRACE simulation run is conducted every time step (steady simulation) in each cell of the computational domain. When the converged perturbed solution is reached, the python script takes charge of setting up the next desired perturbation.

\subsubsection{Discussion about restrictions of the Reynolds stress tensor perturbations}
\label{sec:discussionRestrictions}

\noindent The eigenspace perturbation methodology, presented in \Cref{sec_eigenspaceFramework}, is solely motivated on quantifying the epistemic uncertainties of LEVM due to the inaccurate account for anisotropic flow phenomena. Iaccarino and co-worker \cite{Emory,Iaccarino, MishraSU2} designed this method based on the mathematical derivations of Lumley \cite{Lumley} and Banerjee \cite{Banerjee2007}. These derivations map the states of the Reynolds stress tensor, when it features one, two or three non-zero eigenvalues, onto corners of a constructed triangle, called the barycentric triangle. These states are described to be the extreme states of the Reynolds stress tensor, since the turbulence is only present in one, two or three directions - the corresponding directions of the eigenvectors.\\
A CFD practitioner is interested in ascertain the effect of the turbulence model's uncertainty on certain QoI, which are relevant for design.
However, the relation between the one-, two-, and three-component corner of the Reynolds stress tensor and some QoI is anything but linear. As a consequence, modifying/perturbing the turbulent state of the Reynolds stress tensor seeks to estimate the uncertainty bounds rather than create extreme state of QoI.\\
Nonetheless, we analyzed the relation of barycentric coordinates and QoI for selected flow cases by sampling points inside the barycentric triangle and propagating the perturbed Reynolds stress tensor in an earlier prior investigation \cite{MathaKucharczyk}. Therefore, assessed against currently available data, we agree on the fact, that the corners of the barycentric coordinate produce adequate estimate of the uncertainty bounds in most of the flow regions. Although there might be areas of the flow solution, where the extreme state of turbulence is not corresponding to the extreme state of some QoI. This observation will be also discussed in  \Cref{sec_results}.\\
Additionally, we would like to discuss the effect of the moderation factor $f$, which was initially mentioned by Mishra et al. \cite{MishraSU2}. The main goal of applying $f$ is to reach a converged solution based on the perturbation approach. 
We agree on the fact, that this factor is needed for convergence issues, since some perturbed states tend to be unstable. Understandably, this is especially the case for perturbations seeking to decrease the turbulent production term ($P_{k_\textrm{min}}$ and/or $\mathrm{3C}$). Nevertheless, it is shown in the \nameref{sec:appendix}, that the effect of applying the moderation factor is actually identical to reducing $\Delta_B$ in case of pure eigenvalue perturbation. The need for moderating the effect of Reynolds stress tensor perturbation by an additional factor according to Equation \eqref{reconstructionReStress} emerges, when combining eigenvector and eigenvalue perturbation. Generally speaking, using $f\leq 1$ stabilizes the CFD-simulation by weakening the impact of perturbation.
Accordingly, users are encouraged to not only state the prescribed $\Delta_B$ but also the factor $f$.
It has to be stated, though, that damping the effects of the actual perturbation (eigenvalues and eigenvectors) weakens the interpretability of the limiting states of turbulence, represented by the corners of the barycentric triangle.\\
Last but not least, the data-driven extension of the eigenspace perturbation framework is build on Reynolds stress reference data, whereas other machine learning approaches in the field of turbulence modeling utilize indirect mean flow quantities like velocity and pressure. Hence, appropriate training data is mainly limited to well-resolved DNS/LES and cannot make use of experimental measurement data containing no second-moment statistics.

\subsection{Data sets for training, testing and applying the machine learning model}
\label{sec_dataSets}
\noindent As already described in \Cref{introduction}, the final well-trained machine learning model should be sensitive to flow phenomena such as adverse-pressure gradient, separation and reattachment due to the known shortcomings of the LEVM. 
Consequently, it is reasonable to use data sets for training, which include these flow situations.
We are continuously striving for extending our database, which contains various flow cases for machine learning. For the present study, we use the following flow cases:
\begin{itemize}
    \item DNS of turbulent channel flow at $Re_\tau \in \{180, 550, 1000, 2000, 5200\}$ based on Lee and Moser \cite{lee_moser_2015}
    \item DNS at $Re_H \in \{2800, 5600\}$ and LES at $Re_H = 10595$ of periodic hill flow based on Breuer et al. \cite{Breuer2009}
    \item DNS of wavy wall flow at $Re_H = 6850$ based on Rossi \cite{Rossi2006}
    \item DNS of converging-diverging channel flow at $Re_\tau=617$ based on Laval and Marquillie \cite{Laval}.
\end{itemize}

\noindent All the DNS and LES data of the described test cases are generated using incompressible solvers. In order to simulate these incompressible flows using the compressible solver TRACE without low-Mach preconditioning, the simulations are scaled (adapting dimensions of the geometry and/or molecular viscosity) to an incompressible Mach number of approximately 0.1, while preserving the intended Reynolds numbers. The two-equation, linear eddy viscosity Menter SST $k-\omega$ turbulence model is selected as the baseline model for all conducted RANS simulations \cite{Menter}. In order to obtain steady state solutions, an implicit time marching algorithm is applied. A flux difference splitting approach is employed to discretize the convection terms making use of a second order accurate Roe scheme in combination with MUSCL extrapolation.\\
To evaluate proper features as input parameters based on the RANS simulations, we conducted a mesh convergence study for each of the listed flow cases. Although the mesh convergence studies are not presented here due to the scope of the paper, we affirm, that we only use RANS simulation data, which shows sufficient grid convergence using a low-Reynolds resolution ($y^+ \leq 1$) at solid walls.
The perturbation magnitude $p$ can be determined accordingly to Equation \eqref{eq:perturbationMagnitudeP} by comparing scale-resolving and RANS solutions. In order to compute the intended target quantity, the scale-resolving data has to be interpolated onto the RANS data points for every test case first.
Due to numerical issues some RANS data samples may be located outside the barycentric triangle in terms of anisotropy tensor (see  \Cref{Bild:barycentricTriangle}). Therefore, we included the opportunity to remove these samples from the training or testing sets.
The final application of the UQ perturbation approach is presented for the airfoil test case NACA 4412 at $Re_c = 1.52 \cdot 10^6$.

\subsubsection{Turbulent channel}

\begin{figure}[!h]
\begin{center}
\mbox{\subfigure[Schematic sketch of a fully developed turbulent boundary layer]{\includegraphics[scale=0.45]{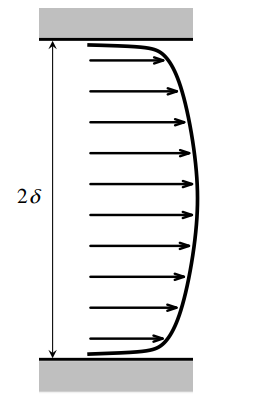}\label{channelFlowSetup1}}}
\mbox{\subfigure[$Re_\tau = 1000$ (every fourth line shown) mesh and boundary conditions; symmetry is enforced in spanwise direction]{\includegraphics[scale=0.45]{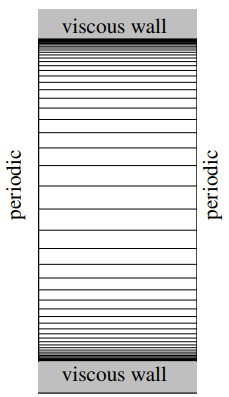}\label{channelFlowSetup2}}}
\vspace{-0.5\baselineskip}
\caption{Turbulent channel flow simulation}
\label{Bild:channelFlow}
\end{center}
\end{figure}

\noindent Although we are interested in more complex cases, the channel flow data serves as one of the key properties, what the model should be able to recognize and predict: turbulent boundary layer with inaccurate anisotropy represented by the LEVM close to the wall. 
The configuration for simulating the turbulent boundary layer is sketched in \Cref{Bild:channelFlow}. The characteristic Reynolds number is defined by
\begin{linenomath}
\begin{equation}
    Re_\tau = \frac{\rho u_\tau \delta}{\mu} \ \text{,}
\end{equation}
\end{linenomath}
where $\delta$ is the channel half-height and the friction velocity is known as $u_\tau = \sqrt{\left(\tau_w/\rho \right)}$ with $\tau_w = \mu \frac{\partial U}{\partial y}|_{y=\textrm{wall}}$.
The turbulent channel flow is homogeneous in the streamwise direction $x$ and the spanwise direction $z$. A constant pressure gradient $\partial P/ \partial x$ is applied to balance the skin friction at the wall.
We use the available RANS grid cells in one half of the channel at the five different Reynolds numbers as subsequent data points for training the random forest.

\subsubsection{Periodic hill}
\noindent The flow over periodic hills features flow separation from curved surfaces, recirculation and a subsequent reattachment on the flat bottom of the channel. Since the Reynolds number has a strong impact on the actual size of the separation bubble, it is worthwhile to add three different Reynolds number flows to our training set.

\begin{figure}[!h]
\begin{center}
\mbox{\subfigure[Relative dimensions and sketch of the flow]{\includegraphics[scale=0.45]{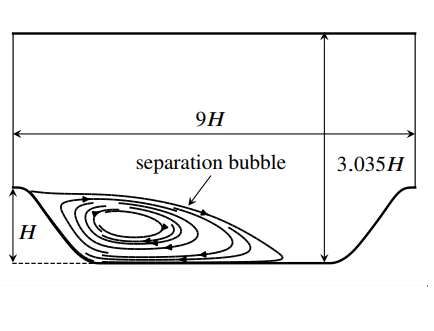}\label{periodicHillSetup1}}}
\mbox{\subfigure[Mesh ($Re_H = 10595$, every tenth line shown) and boundary conditions; slip conditions/inviscid walls are applied in spanwise direction]{\includegraphics[scale=0.45]{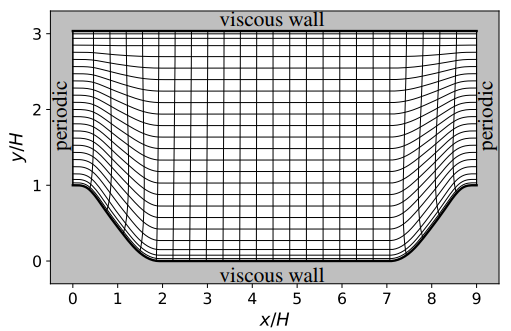}\label{periodicHillSetup2}}}
\vspace{-0.5\baselineskip}
\caption{Schematic periodic hill setup}
\label{Bild:periodicHill}
\end{center}
\end{figure}

The Reynolds number based on the bulk velocity $U_B$, evaluated at the crest of the hill, and the hill height $H$ is defined as
\begin{linenomath}
\begin{equation}
\label{eq:bulkRe}
    Re = \frac{\rho U_B H}{\mu} \ \text{.}
\end{equation}
\end{linenomath}
For simulating the periodic hill configuration, periodic boundary conditions are applied as illustrated in \Cref{Bild:periodicHill}.
A constant pressure gradient $\partial P/ \partial x$ is applied to move the fluid through the configuration. 
The available scale-resolving data sets of the periodic hill only contain data at certain slices ($x/H \in \{0.05, 0.5, 1.0, 2.0,$ $3.0, 4.0, 5.0, 6.0, 7.0, 8.0\}$).
Consequently, the RANS solution is sliced at these locations accordingly. The scale-resolving data is interpolated onto the wall-normal RANS data positions, in order to generate the desired target quantity for the machine learning model.

\subsubsection{Wavy wall}
\begin{figure}[!h]
\begin{center}
\mbox{\subfigure[Relative dimensions and sketch of the flow]{\includegraphics[scale=0.45]{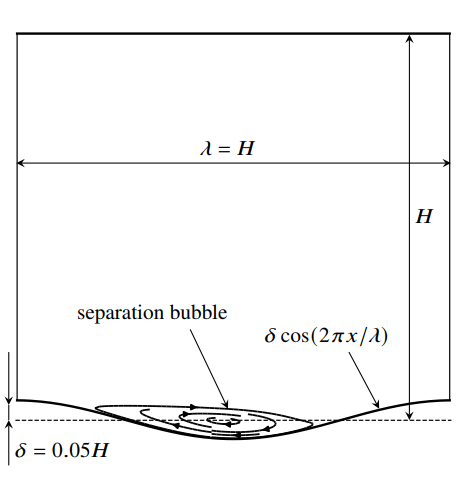}\label{wavyWallSetup1}}}
\mbox{\subfigure[Mesh (every fourth line shown) and boundary conditions; slip conditions/inviscid walls are applied in spanwise direction]{\includegraphics[scale=0.45]{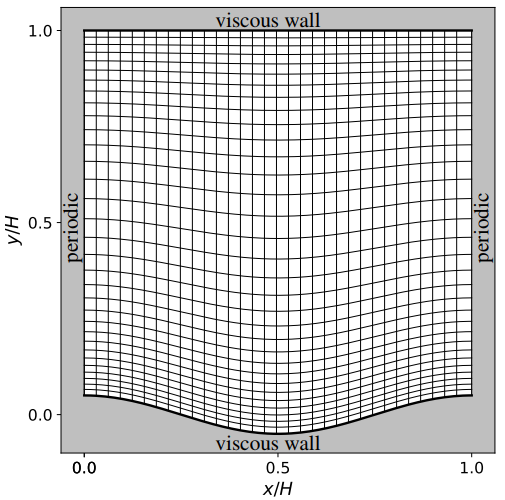}\label{wavyWallSetup2}}}
\vspace{-0.5\baselineskip}
\caption{Schematic wavy wall setup}
\label{Bild:wavyWall}
\end{center}
\end{figure}

\noindent The wavy wall test case is confined by a plane wall and a wavy surface, which is sketched in \Cref{Bild:wavyWall}. 
In former experimental settings, the desired flow situation was generated by stringing together multiple hills and valleys, described by a cosine function. For the CFD simulations (DNS and RANS) periodic boundary conditions in streamwise direction can be applied.
In order to adjust the intended Reynolds number of $Re_H = 6850$, based on the bulk velocity and the mean channel height evaluated on the hill crest (accordingly to Equation \eqref{eq:bulkRe}), a constant pressure gradient $\partial P/ \partial x$ is used. 
Since the available DNS data set is two-dimensional and covers the entire 
domain size, we use all available RANS grid cells as subsequent data points for training the random forest.

\subsubsection{Converging-diverging channel}

\begin{figure}[!h]
\begin{center}
\hspace*{-1cm} 
\mbox{\subfigure[Relative dimensions and sketch of the flow]{\includegraphics[width=1.1\textwidth]{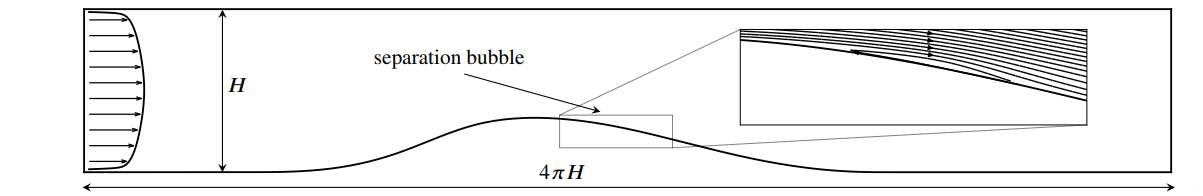}\label{convDivSetup1}}}
\hspace*{-1cm} 
\mbox{\subfigure[Mesh (every fourth line in streamwise direction and every twentieth line in wall normal direction shown) and boundary conditions; slip conditions/inviscid walls are applied in spanwise direction]{\includegraphics[width=1.1\textwidth]{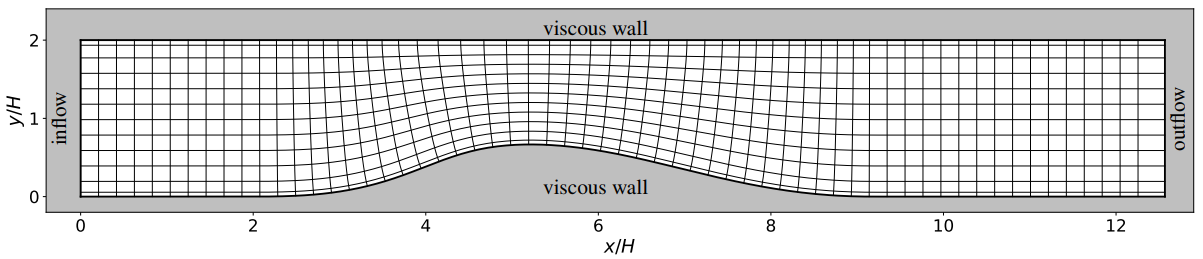}\label{convDivSetup2}}}
\vspace{-0.7\baselineskip}
\caption{Schematic converging-diverging setup}
\label{Bild:convDiv}
\end{center}
\end{figure}

\noindent The configuration of a converging followed by a diverging section is an ideal test case to investigate the effect of an adverse pressure gradient with and without curvature. The flow separates slightly at the diverging part at the lower wall, but not on the flat top wall, as shown in \Cref{Bild:convDiv}.
Similar to the DNS, the inflow boundary conditions are derived from a fully developed turbulent boundary layer at $Re_\tau = 617$ (RANS predicted). A constant mass flow rate is prescribed at the outflow of the domain, which was derived based on the domain size and bulk quantities of the inflow profile (streamwise velocity and density).
Since the available DNS data set is two-dimensional as well, we are able to provide all available RANS grid solution points as subsequent data points for training the machine learning model.

\begin{figure}[!h]
\begin{center}
\mbox{\subfigure[Relative dimensions and sketch of the flow]{\includegraphics[scale=0.45]{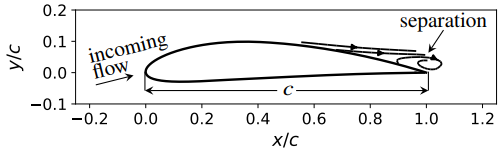}\label{nacaSetup1}}}
\mbox{\subfigure[Mesh (every eighth line shown) and boundary conditions; slip conditions/inviscid walls are applied in spanwise direction]{\includegraphics[scale=0.45]{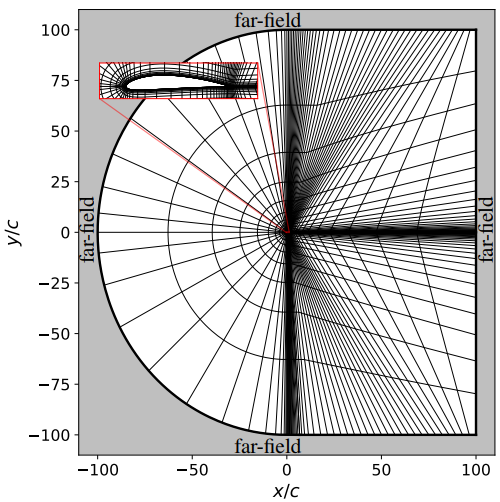}\label{nacaSetup2}}}
\vspace{-0.5\baselineskip}
\caption{Schematic NACA 4412 setup}
\label{Bild:naca4412}
\end{center}
\end{figure}

\subsubsection{NACA 4412 airfoil}
\noindent To demonstrate the application of the UQ framework with and without a machine learning model, the near-stall NACA 4412 airfoil is chosen in the presented work. This test case is a NASA benchmark case for turbulence models, featuring boundary layer separation close to the trailing edge. This airfoil is operated at a Reynolds number of $Re_c = 1.52 \cdot 10^6$ (based on the freestream velocity $U_{\inf}$ and the chord length $c$) and a Mach number of $Ma = 0.09$ (based on $U_{\inf}$). The angle of attack is $13.87^{\circ}$ as sketched in \Cref{nacaSetup1}. The CFD results are compared against experimental measurements of Coles and Wadcock \cite{Wadcock1979}. In order to minimize the effect of boundary conditions on the CFD simulation, far-field conditions are applied to prescribe the specified flow conditions (see \Cref{nacaSetup2}). A turbulence intensity of 0.086\% and an eddy viscosity ratio $\mu_t/\mu$ of 0.009 is prescribed in accordance with the description of NASA \cite{NASATurbulenceModelling}. The mesh topology is the so-called C-grid featuring a grid cell resolution of $n_x, n_y, n_z = 896,256,1$, which can be downloaded from NASA's turbulence database \cite{NASATurbulenceModelling}. 

\section{Hyperparameter selection based on generalization study}
\label{sec_hyperparameterStudy}
\noindent Before the training of the final random forest regression model was conducted, the impact of four different hyperparameters:
\begin{itemize}
    \item maximum tree depth: maximum number of decision nodes from the root down to the furthest node allowed 
    \item minimum sample count: minimum number of data samples required at a decision node allowed
    \item maximum number of features: maximum number of features randomly chosen at each decision node allowed
    \item number of trees: total number of individual decision trees used
\end{itemize} 
on the accuracy is evaluated.
Since the final trained model should be able to generalize for different geometries and flow conditions, it seems to be reasonable to evaluate these hyperparameters with focus on generalization capabilities of the random forest. Therefore, we apply a leave-one-out-cross validation, which is an appropriate procedure for small data sets. This means, that three out of four available training data sets (see \Cref{sec_dataSets}) are used for training, while the remaining flow case is used to verify the model (see \Cref{tab:hyperparameters}). Data samples featuring non-physical Reynolds stress tensors (barycentric coordinates are located outside the barycentric triangle) are removed from each data set.

\begin{table}[htb]
  \caption{Scenarios for hyperparameter study: $x$ means part of training data, $\circ$  means testing data}
  \label{tab:hyperparameters}
  \centering
  {\footnotesize
  \begin{tabular}{lllll}
    \toprule
                       &         \multicolumn{4}{c}{Scenario}                   \\
    \cmidrule(r){2-5}
    Flow cases          & I     & II & III & IIII\\
    \midrule
    turbulent channel                 & x  & x  & x & $\circ$ \\
    periodic hill                     & x  & x  & $\circ$  & x\\
    wavy wall                         & x  & $\circ$   & x & x\\
    converging-diverging channel      & $\circ$   & x  & x & x\\
    \bottomrule
  \end{tabular}
  }
\end{table}

\begin{figure}[!h]
\begin{center}
\mbox{
\subfigure[Maximum depth of trees over total number of individual trees]{\includegraphics[width=0.3\textwidth, trim=0.4cm 0cm 0.3cm 0.2cm, clip=True]{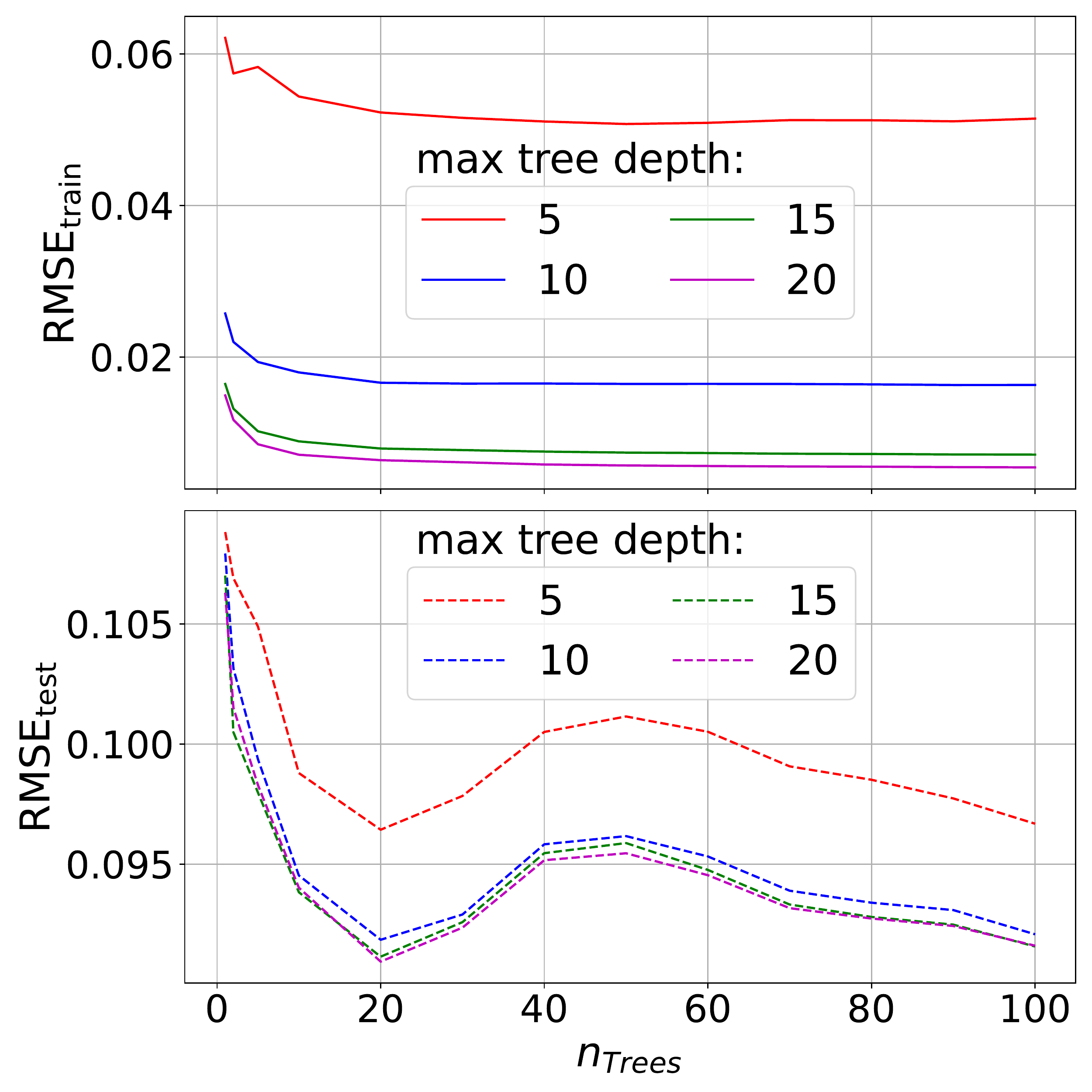}\label{maxDepthCase1}}}
\mbox{
\subfigure[Minimum number of required samples over total number of individual trees]{\includegraphics[width=0.3\textwidth, trim=0.4cm 0cm 0.3cm 0.2cm, clip=True]{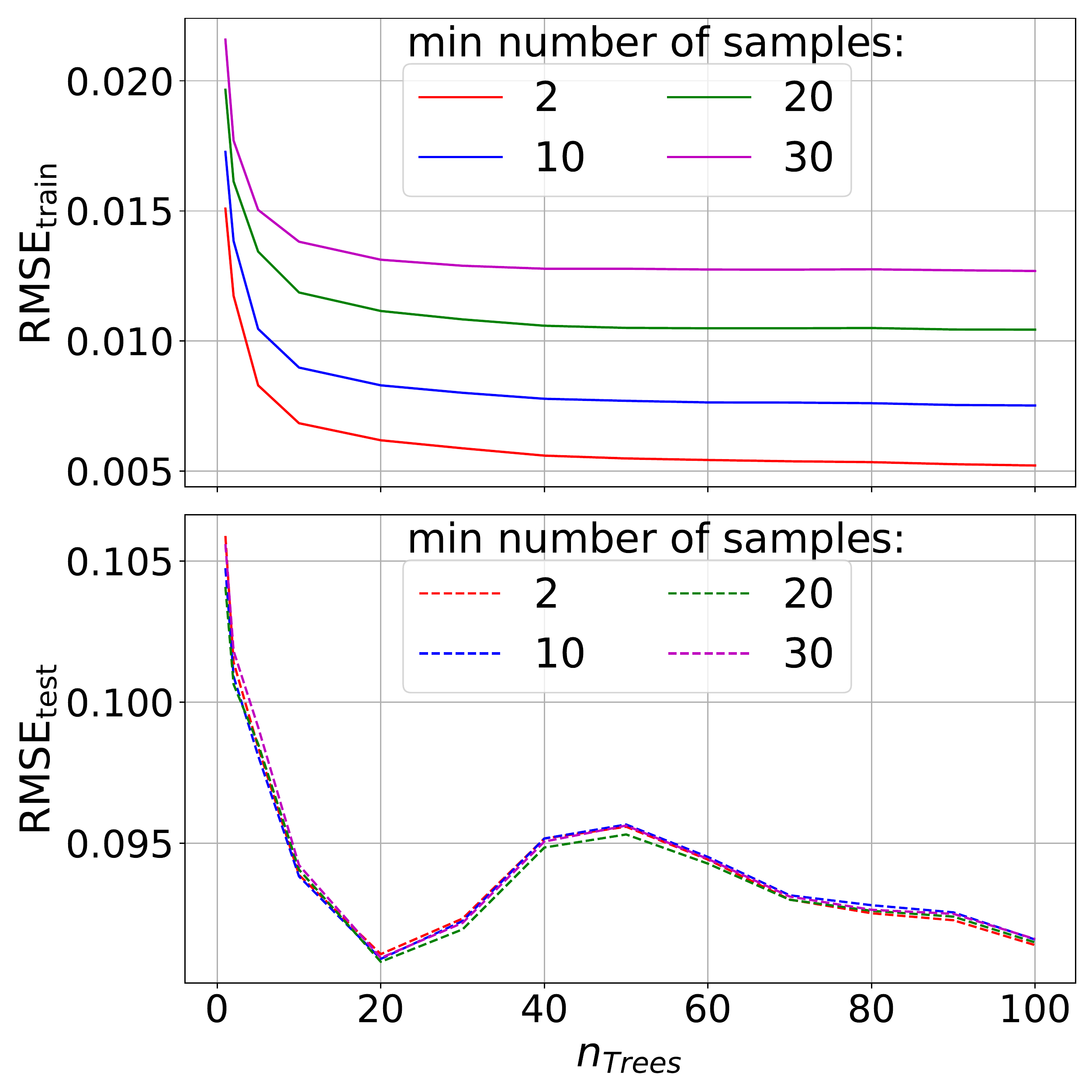}\label{minSamplesSplitCase1}}}
\mbox{
\subfigure[Maximum number of active features over total number of individual trees]{\includegraphics[width=0.3\textwidth, trim=0.4cm 0cm 0.3cm 0.2cm, clip=True]{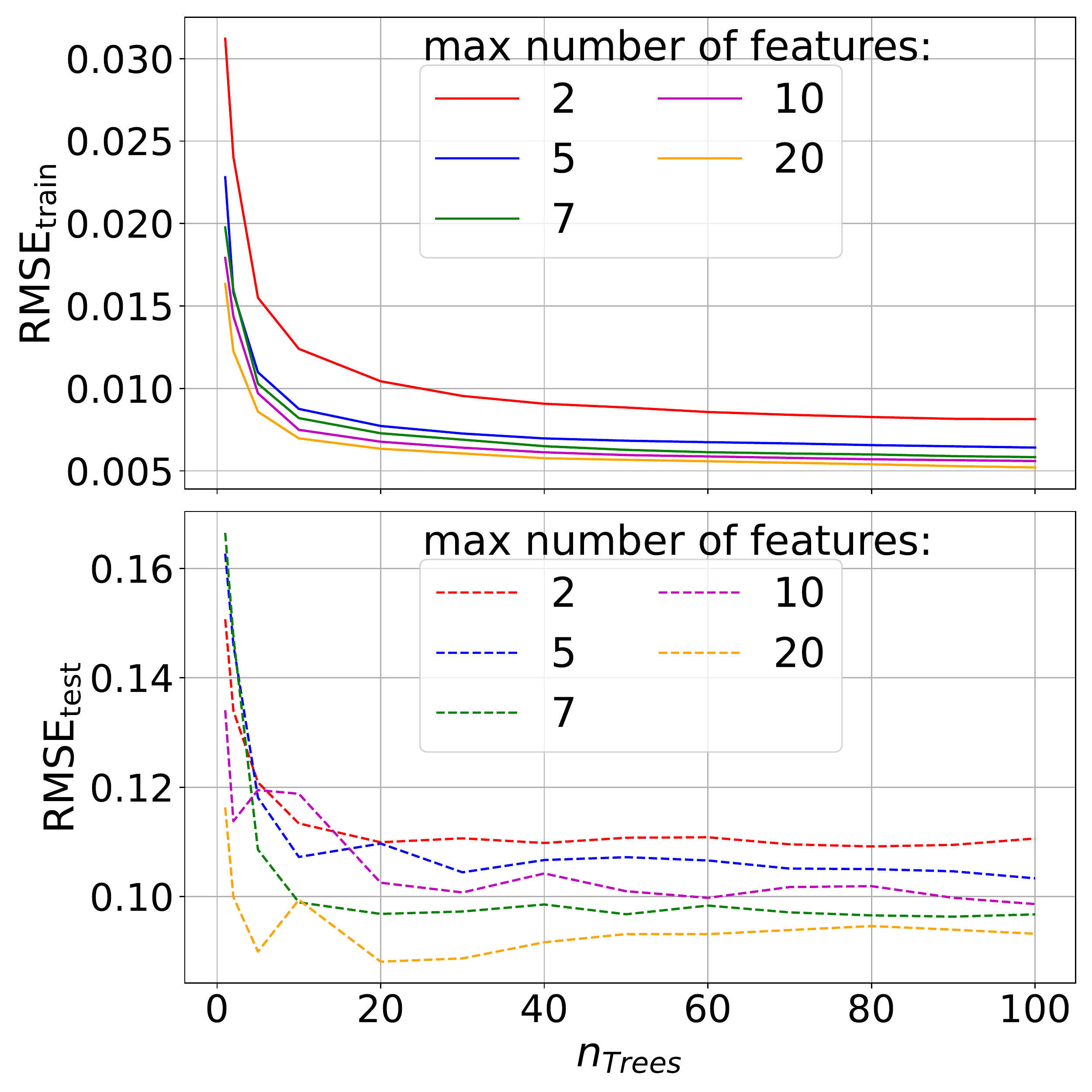}\label{maxFeaturesCase1}}}
\caption{Training accuracy (solid / top) and testing accuracy (dashed / bottom) based on $\mathrm{RMSE}$ for selection of hyperparameters in scenario I}
\label{Bild:hyperparameters}
\end{center}
\end{figure}

\noindent For each of the first three hyperparameters several different values were studied over a range of the total amount of individual regression trees, while the other two hyperparameters were set to default values (see \textit{scikit-learn} documentation for further information).
As an example, \Cref{Bild:hyperparameters} presents the effect of the considered hyperparameters on the accuracy of the model prediction in scenario I, where the accuracy is expressed in terms of the root mean square error ($\mathrm{RMSE}$).

\noindent The subsequent statements made coincide with all evaluated scenarios as well.
With increasing maximum depth of each individual tree, the model prediction error on the training data is reduced at the expense of increasing complexity. Increasing complexity of the tree may result in reduced generalizability capabilities (reduced test data accuracy) of the  model. Because a maximum tree depth of 15 showed excellent performance for the training as well as for testing data, it is preferred compared to a higher value of 20, which increases computational costs (see \Cref{maxDepthCase1}).
The smaller the number of data samples for each decision node, the more accurate the performance on the training data (see \Cref{minSamplesSplitCase1}).
Since the $\mathrm{RMSE}$ based on evaluated testing data is not significantly affected by this hyperparameter, a minimum sample count of 10 is chosen. This enables the model to generalize to a greater extend, than selecting a smaller value.
A larger number of selected features for each decision node lowers the training error and increases the risk of overfitting. Since a maximum number of 7 features produces accurate prediction performance for the test data as well (see \Cref{maxFeaturesCase1}), it is selected as inferred hyperparameter.
In terms of total number of individual trees, one can observe a steep drop in $\mathrm{RMSE}$ for small numbers followed by a constant level of accuracy. The computational costs scale linearly with the number of individual trees.
Although computational costs do not really play a relevant role for our application, as we only evaluate the model once before each simulation run, we sought for the minimum number of trees for maximized performance of the model. Therefore, we concluded to use a total number of 30 individual trees for the random forest by evaluating all described scenarios in \Cref{tab:hyperparameters}.

\section{Verification of trained machine learning model}
\label{sec_Modelvalidation}

\noindent Based on the choice of hyperparameters, which was discussed in the previous section, the prediction accuracy of the random forest should be evaluated on the available data (see \Cref{sec_dataSets}). Ten different scenarios based on combinations of the flow data cases, listed in \Cref{tab:predictionAccuracy}, serve as verification of functionality and present the accomplishment of the intended generalization of every model. 
While data samples featuring non-physical Reynolds stress tensors (barycentric coordinates are located outside the barycentric triangle) are removed from the data sets, the $\mathrm{RMSE}$ classifies the prediction accuracy of different scenarios. As the target quantity $p$ can vary between zero and one (based on the construction of the equilateral triangle with edge length equals to one), the resulting $\mathrm{RMSE}$ indicate less than 10\% absolute prediction error except for scenario II. 

\begin{table}[!t]
  \caption{Prediction accuracy of random forest: $x$ means part of training data, $\circ$ means \underline{not} part of training data, \textcolor{red}{red} means data set for evaluation of $\mathrm{RMSE}$}
  \label{tab:predictionAccuracy}
  \hspace*{-1.2cm}
  \centering
  {\footnotesize
  \begin{tabular}{llllllllllll}
    \toprule
                       &  &       \multicolumn{10}{c}{Scenario}                   \\
    \cmidrule(r){3-12}
    \multicolumn{2}{l}{Flow cases}          & I     & Ia & II & IIa & III & IIIa & IIIb & IIII & IIIIa& V\\
    \midrule
    \multicolumn{2}{l}{turbulent channel}                 & x                        & x                   & x                          & x                  & x                         & x            &x     &\textcolor{red}{$\circ$} & \textcolor{red}{x}&\textcolor{red}{x}\\
    \multicolumn{2}{l}{periodic hill}                     &&&&&&&&&\\
    & \footnotesize{$\cdot \ Re_H \in \{2800, 5600, 10595\}$}  & x                        & x                   & x                          & x                  &\textcolor{red}{$\circ$}   &\textcolor{red}{x} & &x&x&\textcolor{red}{x}\\
    &\footnotesize{$\cdot \ Re_H \in \{2800, 10595\}$}  &   &  & &  &   & &x &&\\
    &\footnotesize{$\cdot \ Re_H = 5600$}  &   &  & &  &   & &\textcolor{red}{$\circ$} &&&\\
    \multicolumn{2}{l}{wavy wall}                         & x                        & x                   & \textcolor{red}{$\circ$}    & \textcolor{red}{x} &x                          &x&x &x&x&\textcolor{red}{x}\\
    \multicolumn{2}{l}{converging-diverging channel}      & \textcolor{red}{$\circ$} & \textcolor{red}{x}  & x                          & x                  &x                          &x&x &x&x&\textcolor{red}{x}\\
    \midrule
    \multicolumn{2}{l}{$\mathrm{RMSE}\left(p_{\mathrm{pred}},p_{\mathrm{true}}\right)$} & 0.098 & 0.010 & 0.133 & 0.029&  0.095&0.028& 0.041& 0.051& 0.014 & 0.013     \\
    \bottomrule
  \end{tabular}
  }
\end{table}

\noindent As soon as a trained machine learning model should make predictions on flow cases, for which accurate data does not exist, judging the model's prediction in terms of accuracy becomes difficult.
Comparing the input feature spaces of training and testing data (previously unseen case) resulting in extrapolation metrics, in order to build confidence in a machine learning model is a reasonable idea.
Extrapolation metric measures the distance between a test point $\tilde{m}$ and the training data feature set $m^{(i)}$ for $i = 1,\dots,n$ with $n$ as the number of training data points.
In this paper, we use the Kernel Density Estimation (KDE) to compute the distance by estimating the probability density
\begin{linenomath}
\begin{equation}
\label{eq:kernel}
    f_{\mathrm{KDE}} = \frac{1}{n \sigma^{d}} \sum_{i=1}^n \prod_{j=1}^d K \left(\frac{\tilde{m}_j-m^{(i)}_{j}}{\sigma}\right) \ \text{,}
\end{equation}
\end{linenomath}
with the number of features $d$ and the bandwidth $\sigma$, determined by Scott's rule \cite{Scott}. According to the work of Wu et al. \cite{Wu}, we use a Gaussian kernel $K\left(t\right) = 1/\sqrt{\left(2 \pi \right)} \exp{\left(-t^2/2\right)}$ and the distance is computed as follows:
\begin{linenomath}
\begin{align}
\label{eq:kdeDistance}
    \begin{split}
    d_{\mathrm{KDE}} &= 1-\frac{f_{\mathrm{KDE}}}{f_{\mathrm{KDE}}+1/A} \quad \text{with} \\
    A&=\prod_i^n \left(\max_j\left({m^{(i)}_{j}}\right) - \min_j\left({m^{(i)}_{j}}\right)\right) \quad \text{for} \quad j = 1,\dots,d
    \end{split}
\end{align}
\end{linenomath}
The applied Gaussian kernel $K$ ensures, that the smaller the difference of $\tilde{m}_j$ and $m^{(i)}_{j}$, the larger the output of $K$. In other words, Equation \eqref{eq:kernel} increases, if $\tilde{m}$ becomes close to a concentrated feature space of the training data points and vice versa. The quantity $A$, which is only dependent on the features of the training set, can be interpreted as the volume of a cuboid with $d$ dimensions. Thus $1/A$ is the probability density with respect to a uniform distribution inside such a cuboid.
Due to the normalization of the distance in Equation \eqref{eq:kdeDistance}, the metric is able to measure the distance of $\tilde{m}$ to the training data with respect to a uniform distribution.
Consequently, on the one hand, if $\tilde{m}$ is close to a concentrated feature space, $f_{KDE} \gg 1/A$ implies $d_{\mathrm{KDE}}\rightarrow 0$. 
On the other hand, $d_{\mathrm{KDE}}\rightarrow 1$ follows from $f_{KDE} \ll 1/A$.
This enables a user to interpret the rate of extrapolation needed based on the training data set. 
Thus, two extreme scenarios are represented according to:
\begin{itemize}
    \item $d_{\mathrm{KDE}}=0$: no extrapolation is required; the features of the training data set cover the feature of the test point $\tilde{m}$
    \item $d_{\mathrm{KDE}}=1$: high extrapolation is required; the features of the test point $\tilde{m}$ are far off the features of the training data.
\end{itemize}

\begin{figure}[!h]
\begin{center}
\includegraphics[scale=0.33, trim=0.5cm 0cm 0.7cm 0.1cm, clip=False]{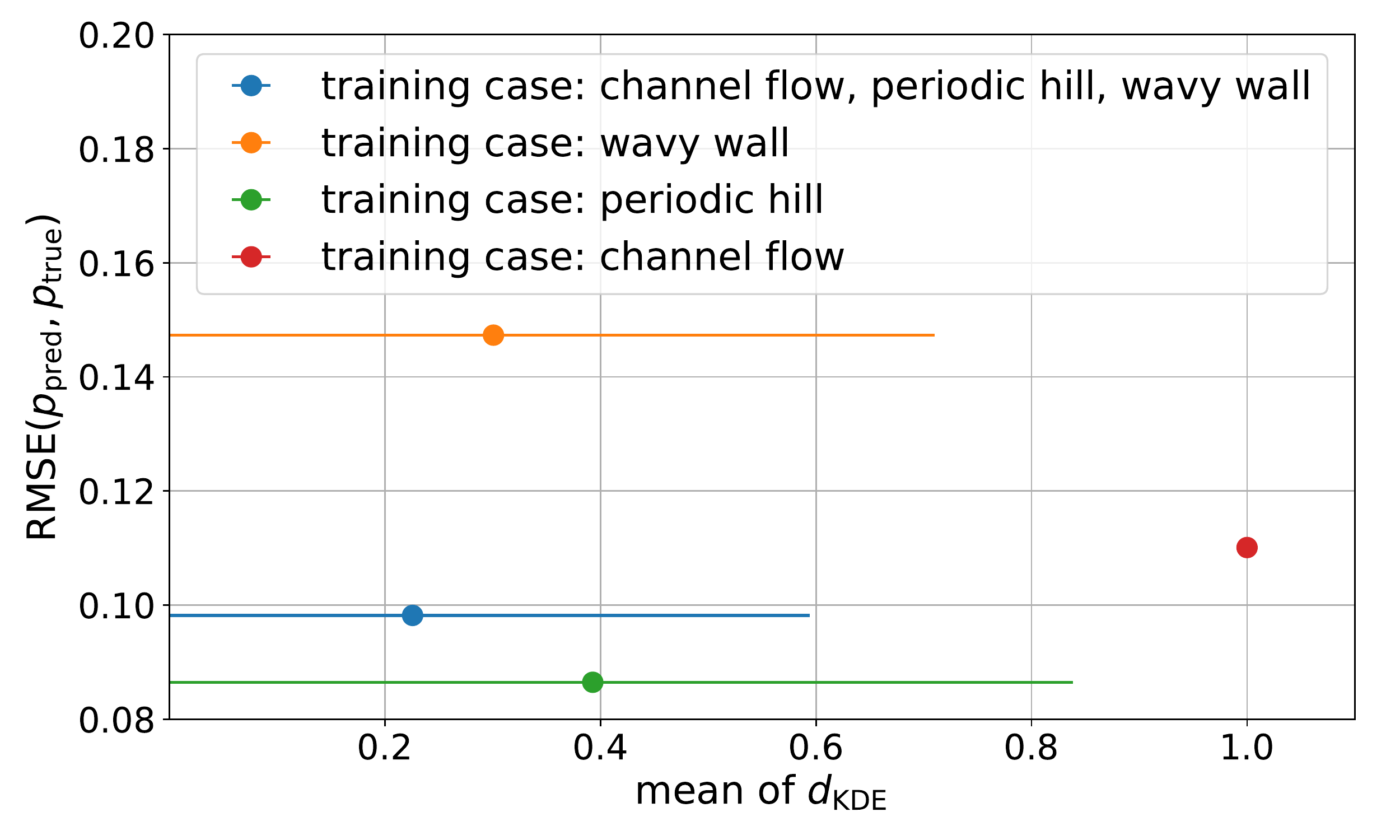}
\vspace{-0.5\baselineskip}
\caption{Relationship between the RMSE of the prediction for the converging-diverging channel and the mean value of KDE extrapolation metric (standard deviation of the extrapolation metric is shown as the horizontal bars). All 56 input features are considered for the prediction and training of the random forest models, while only $q_1$, $q_2$, $q_3$, $q_7$ and $q_8$ are used to compute $d_{\mathrm{KDE}}$.}
\label{Bild:convDivCorrelation}
\end{center}
\end{figure}

Since the extrapolation metric only assesses the closeness of the features between training and test data sets, Wu et al. \cite{Wu} demonstrate, that the KDE extrapolation metric can be used to estimate the prediction confidence by quantifying the correlation between the degree of extrapolation and the prediction accuracy.
In our work, the flow case of the converging-diverging channel serves to present the application of the extrapolation metric. The remaining data (turbulent channel flow, flow over periodic hills and wavy walls) is used for training individual random forests, while each random forest is evaluated on the converging-diverging channel.\\
Selected input features attributed with significant feature importance are considered for computing the KDE distance $d_{\mathrm{KDE}}$ .
The individual feature importance for each of the 56 input features, is determined after training the final random forest (scenario V in \Cref{tab:predictionAccuracy}) using the chosen hyperparameters (see \Cref{sec_hyperparameterStudy}). The utility of each feature is determined by the permutation feature importance approach, accounting for the reduction in the model accuracy, when the values of this feature are randomly shuffled.   
Consequently, the selected five most important features to be considered for determining $d_{\mathrm{KDE}}$ are the eddy viscosity $q_8$, the normalized wall-distance $q_3$, the Mach number $q_7$, the turbulent kinetic energy $q_2$ and the Q-criterion $q_1$.
Contrary to the work of Wu et al. \cite{Wu}, we cannot confirm a strong correlation between the accuracy of the model, evaluated by predicting the perturbation magnitude $p$ for the converging-diverging channel, and the mean of the KDE distance $d_{\mathrm{KDE}}$ for different training data sets, as illustrated in \Cref{Bild:convDivCorrelation}.

\begin{figure}[!t]
\begin{center}
\mbox{\subfigure[Converging-diverging wall ]{\includegraphics[scale=0.43]{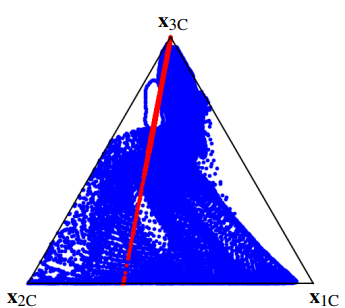}\label{convDivbary}}}
\mbox{\subfigure[Periodic hill, $Re_H = 5600$ ]{\includegraphics[scale=0.43]{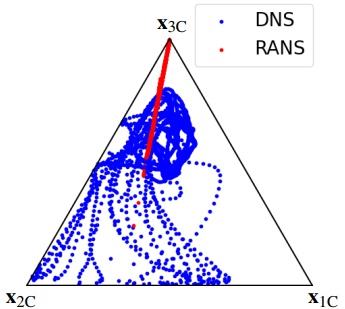}\label{hillbary}}}
\mbox{\subfigure[Wavy wall ]{\includegraphics[scale=0.43]{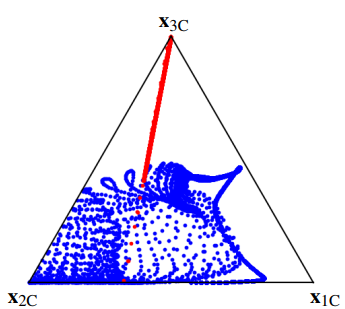}\label{hillWavy}}}
\vspace{-0.5\baselineskip}
\caption{Barycentric coordinates for selected flow cases; legend of (b) corresponds to (a) and (c) as well}
\label{Bild:targetBaryComparison}
\end{center}
\end{figure}

\noindent However, \Cref{Bild:targetBaryComparison} presents a possible explanation for reduced prediction error, when training on the periodic hill compared to the wavy wall. The DNS data based barycentric coordinates of the converging-diverging channel and the periodic hill cover similar areas in the barycentric triangle, while true values of barycentric coordinates for the wavy wall test case are only located in the lower range of the triangle. 

\begin{figure}[!t]
\begin{center}
\mbox{
\subfigure[Converging-diverging channel ]{\includegraphics[scale=0.143]{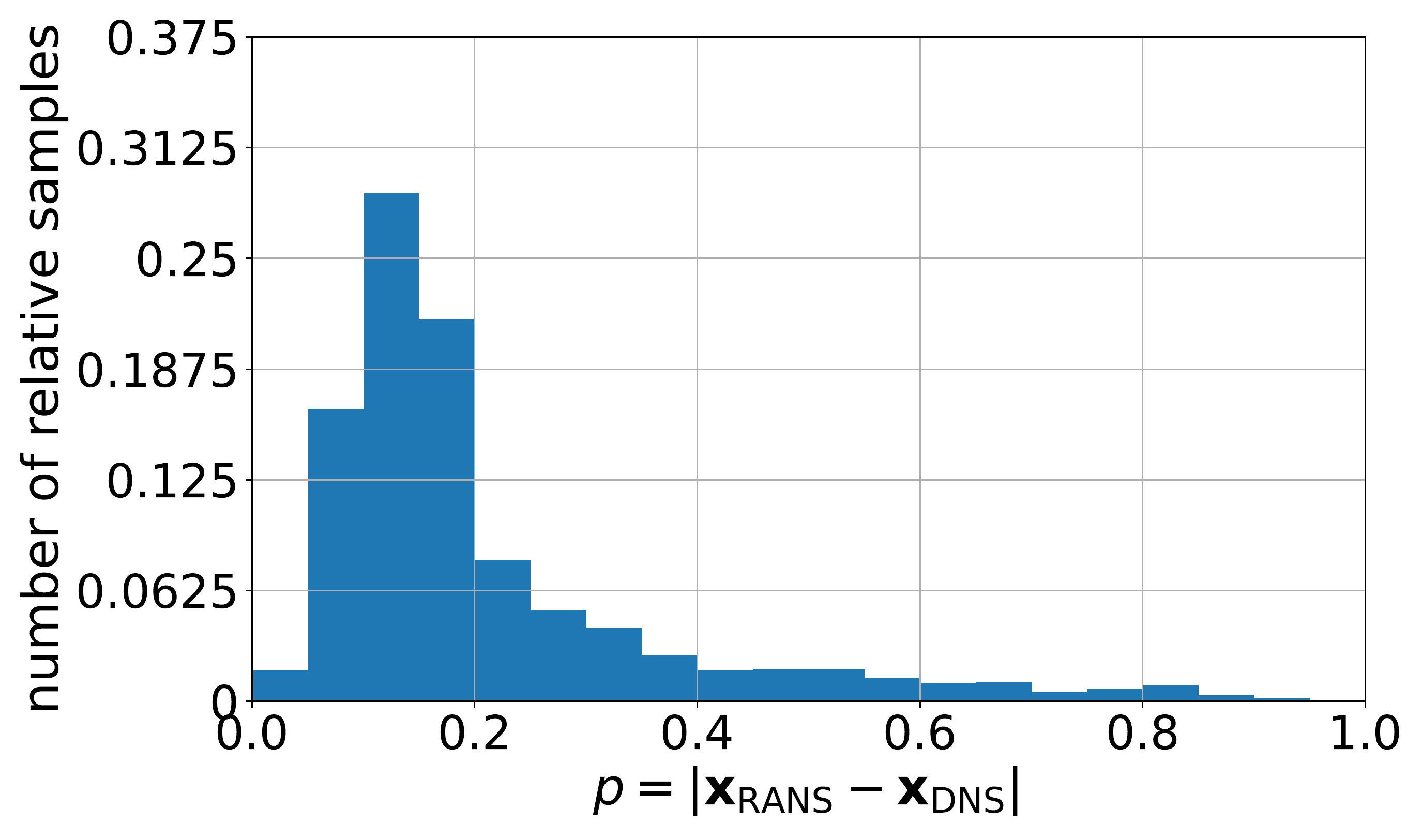}\label{histoconvDivbary}}}
\mbox{
\subfigure[Periodic hill, $Re_H = 5600$]{\includegraphics[scale=0.14, trim=0.1cm 0cm 0cm 0cm, clip=true]{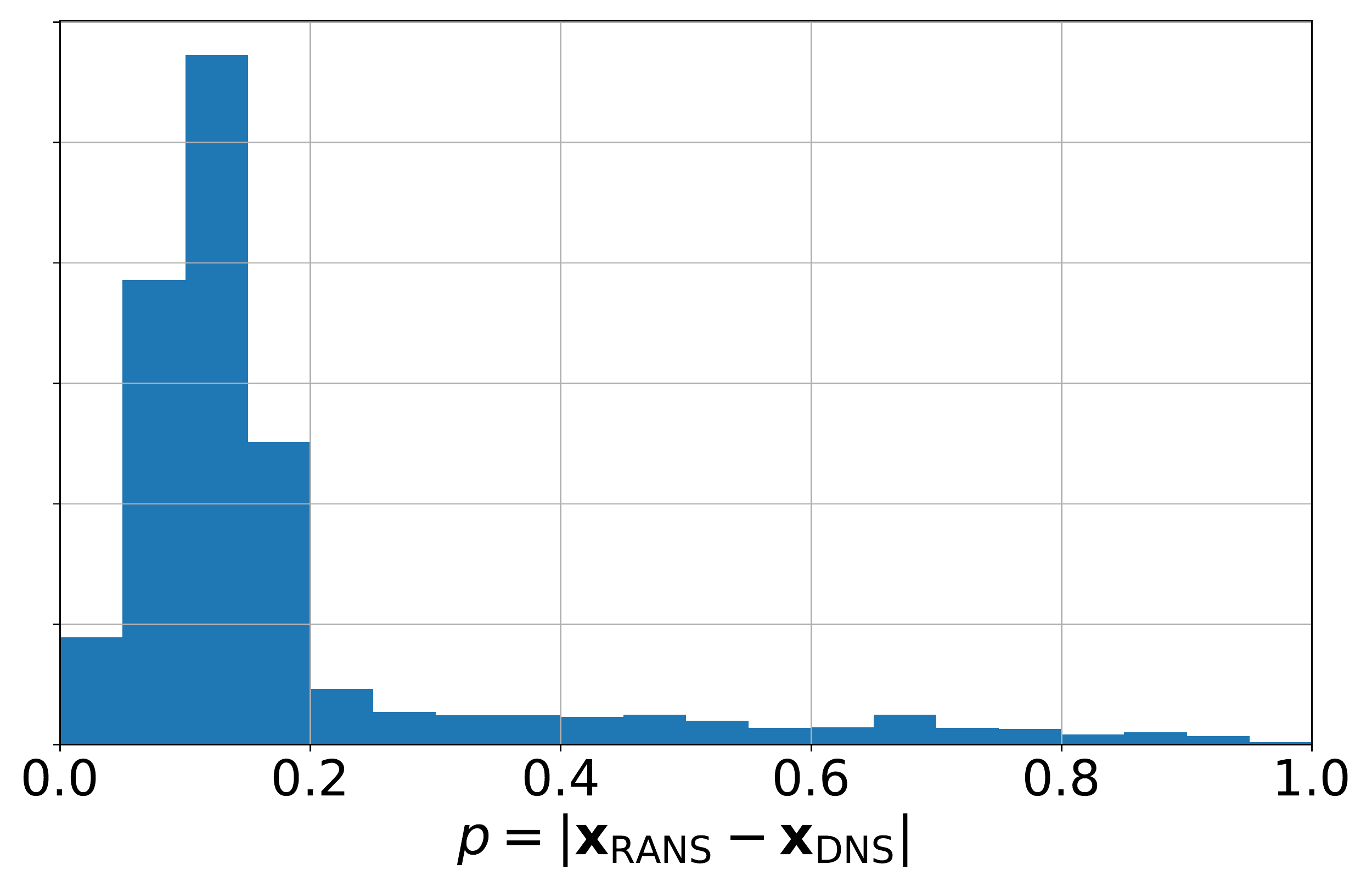}\label{histohillbary}}}
\mbox{
\subfigure[Wavy wall]{\includegraphics[scale=0.14, trim=0.1cm 0cm 0cm 0cm, clip=true]{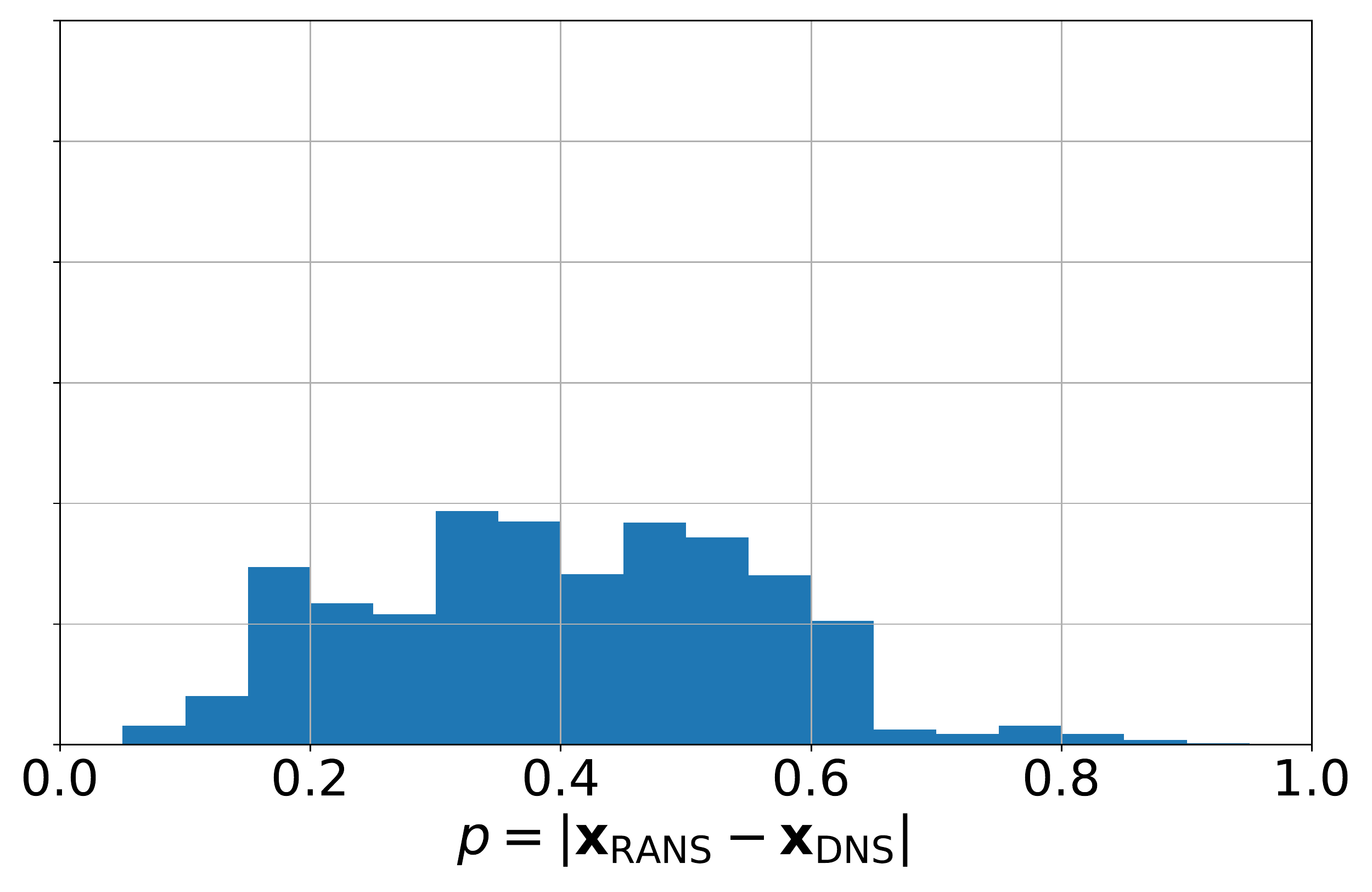}\label{histohillWavy}}}
\vspace{-0.5\baselineskip}
\caption{Frequency of target quantity $p$ for selected flow cases; vertical axis correspond to (a), (b) and (c)}
\label{Bild:histogramBaryComparison}
\end{center}
\end{figure}

\noindent Thus, the target quantity, which is the distance in barycentric coordinates, becomes more frequent in a similar range of absolute values for the converging-diverging channel and the periodic hill (see \Cref{Bild:histogramBaryComparison}). Even Wu et al. \cite{Wu} mention, that the correlation between accuracy and extrapolation metric is less correlated, if the training set is very similar or very different from the test set, what we might be facing here as well.

Nevertheless, the result of the extrapolation metric is highly dependent on the set of considered features. Thus, it seems reasonable to limit the evaluation of the metric to certain important features for the random forest.

\begin{figure}[!t]
\begin{center}
\hspace*{-0.55cm}
\mbox{
\subfigure[Evaluated extrapolation metric]{\includegraphics[width=1.15\textwidth, trim=0.5cm 0cm 0.7cm 0.1cm, clip=False]{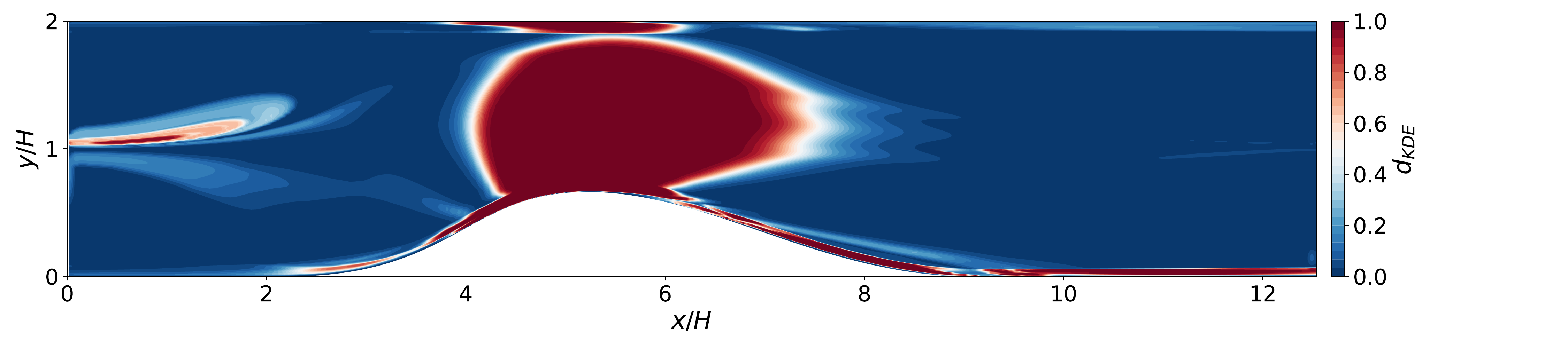}\label{convDivMetric}}}
\hspace*{-0.65cm}
\mbox{
\subfigure[Model prediction $p_\mathrm{pred}$ presented as absolute deviation based on true values]{\includegraphics[width=1.18\textwidth]{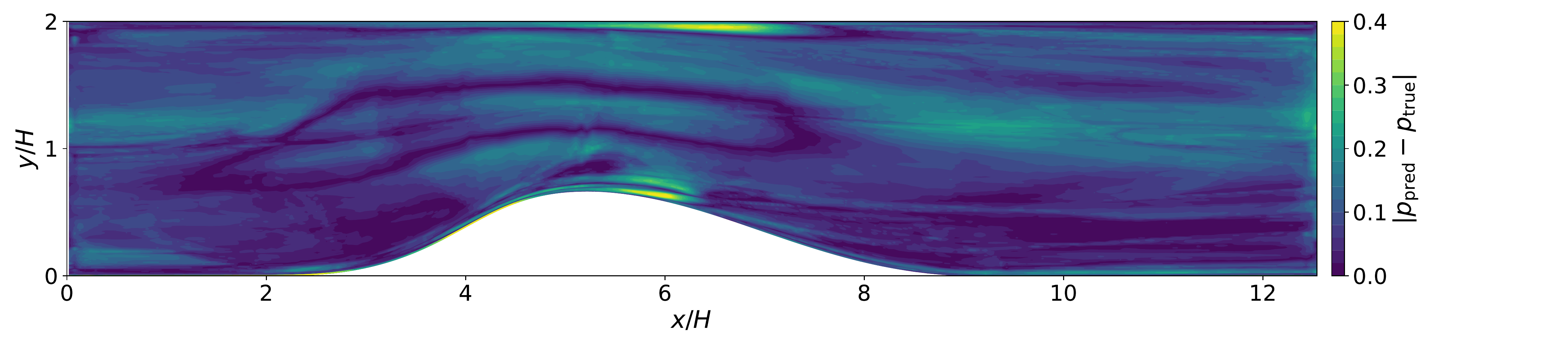}\label{convDivP}}}
\vspace{-0.5\baselineskip}
\caption{Verification of metric for converging-diverging channel}
\label{Bild:convDivMetric}
\end{center}
\end{figure}

\Cref{convDivMetric} presents the two-dimensional distribution of the KDE metric, evaluated based on the five most important features $q_8, q_3, q_7, q_2$ and $q_1$ (corresponding to the blue data point in  \Cref{Bild:convDivCorrelation} and scenario I in \Cref{tab:predictionAccuracy}).
Although, some spatial correlated regions between KDE distances $d_{\mathrm{KDE}}$ and model errors $|p_{\mathrm{pred}}-p_{\mathrm{true}}|$ in \Cref{convDivP} can be recognized, their overall correlation is not strong (Pearson correlation coefficient of approx. 0.2). Nonetheless, similar observations based on comparable test cases as in the work of Wu et al. \cite{Wu} (not shown here) reinforce the trust in the presented KDE distance, when applied in an adequate manner. 

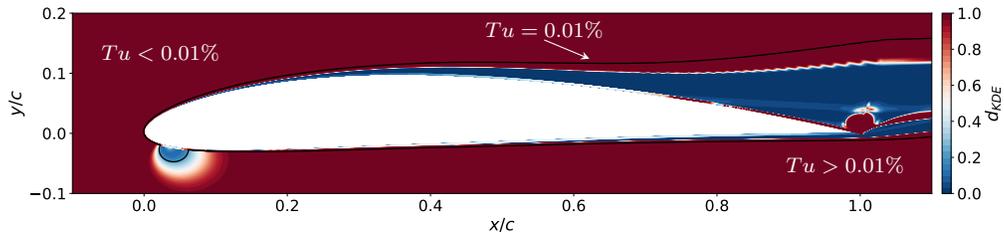
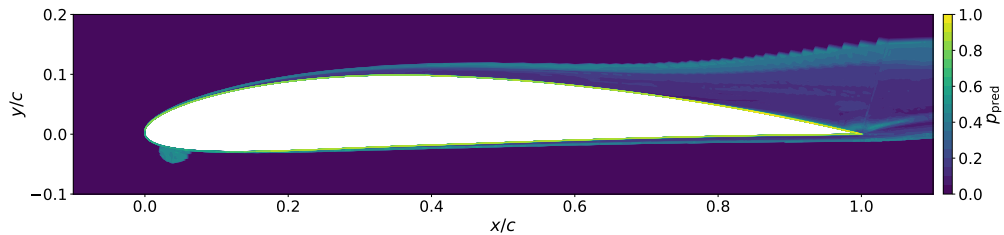
\begin{figure}[b!]
\begin{center}
\hspace*{-1.05cm}
\begin{tikzpicture}
\node[] at (0,0) {\mbox{
\subfigure[Evaluated extrapolation metric, black line divides areas with higher and lower turbulence intensity $Tu$ with a threshold of $0.01\%$]{\includegraphics[scale=0.322, trim=0.5cm 0cm 0.7cm 0.1cm, clip=False]{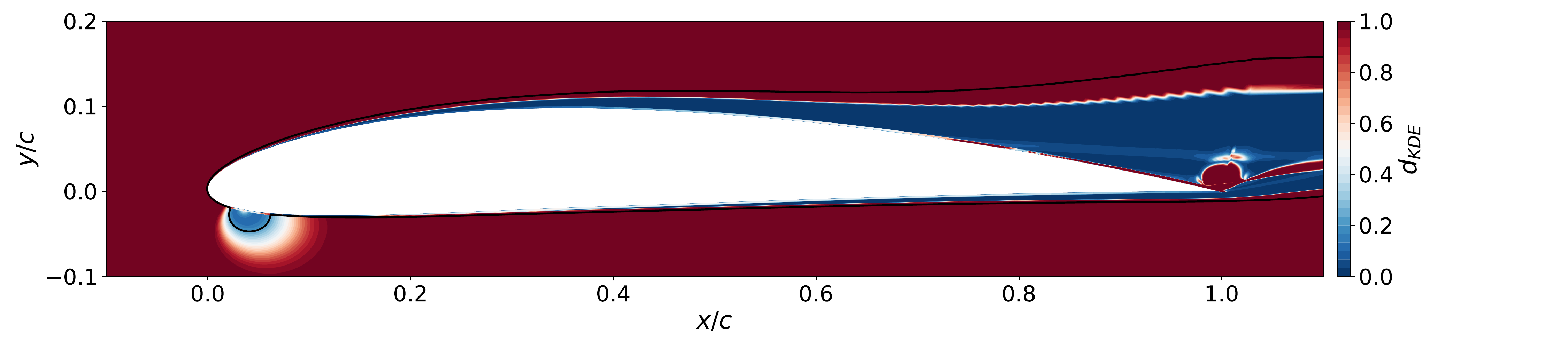}\label{Naca4412Metric}}}};
\node[white] at (-5.1, 1.5) {\footnotesize $Tu < 0.01 \%$};
\node[white] at (0, 1.8) {\footnotesize $Tu = 0.01 \%$};
\draw [white, -stealth](0,1.67) -- (0.6,1.42) node[] {};
\node[white] at (4, 1-1) {\footnotesize $Tu > 0.01 \%$};
\end{tikzpicture}
\hspace*{-1.05cm}
\mbox{
\subfigure[Model prediction in the area of $Tu \geq 0.01 \%$ ($p_\mathrm{pred}$ is set to zero for area with $Tu < 0.01 \%$)]{\includegraphics[scale=0.322]{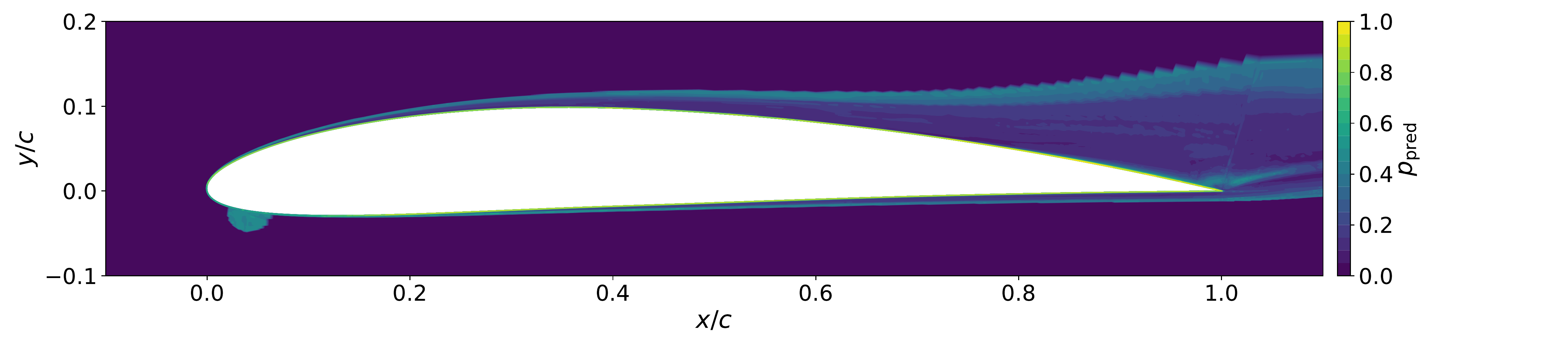}\label{Naca4412P}}}
\vspace{-1\baselineskip}
\caption{Evaluated metric and perturbation magnitude for NACA 4412 profile}
\label{Bild:NACA4412Metric}
\end{center}
\end{figure}

Before the actual application of the data-driven UQ perturbation framework on the flow around NACA 4412 can be conducted, the prediction of the random forest model for this case should be discussed. \Cref{Bild:NACA4412Metric} shows the evaluated extrapolation metric and the corresponding random forest predicted $p$.
The estimated distribution of the KDE distance based on the five most important flow features $q_8, q_3, q_7, q_2$ and $q_1$ is shown in \Cref{Naca4412Metric}. Solely regions close around the airfoil contain $d_{\mathrm{KDE}}\leq 1$. This is due to the widely differing geometry, Reynolds number and flow situation compared to the training data. The feature space of NACA 4412 contains especially higher Mach numbers $q7$ in regions far off the boundary layer with less turbulent kinetic energy $q2$ (respective turbulent eddy viscosity $q8$) and limited wall-distance based Reynolds Number $q3$. This fact also manifests, when taking a look at the level of turbulence intensity for the baseline solution. Because far-field boundary conditions are used, laminar flow is present almost everywhere in the CFD domain. An area with $Tu \geq 0.01\%$ can only be identified in the boundary layer around the profile, in the wake flow and around the stagnation point on the pressure side at $x/c \approx 0.035$. Consequently, the chosen training data does not sufficiently cover the feature space of NACA 4412 in the outer regions. But since the entire mapping of the Reynolds stress tensor onto barycentric coordinates makes only sense for relevant turbulent stresses, which are significantly larger than machine precision, the random forest is restricted to predict values in the area at $Tu \geq 0.01\%$ (see \Cref{Naca4412P}). This decision is purely based on our observations and physical intuition.
The verification of this procedure was numerically justified by comparing forward data-driven UQ computations based on model predicted $p$. Some simulations contained $p$, determined by the random forest, everywhere in the domain, others only in the area $Tu \geq 0.01\%$. Although the random forest is able to predict certain values greater than zero in the region featuring $Tu \leq 0.01\%$, the evaluated flow quantities around the NACA 4412 airfoil did not show any significant difference (not shown here). 
Even the KDE distance in \Cref{Naca4412Metric} confirms the chosen procedure, by revealing reduced extrapolation distance in areas with increased turbulence intensity.
Thus, restricting the model prediction to certain area closely around the airfoil can be justified from a physical and a machine learning perspective. However, the presented extrapolation metric and model prediction close to the separation zone ($x/c \geq 0.8$ based on RANS simulation) reveal, that predicting the influence of a separated region in terms of anisotropy discrepancy is a challenging task. This issue might be only overcome with an increasing number of training data sets involving varying flow conditions and geometries.

\section{Application of UQ perturbation framework}
\label{sec_results}
\begin{figure}[!t]
\begin{center}
\mbox{
\subfigure[Pressure coefficient $c_p$]{\includegraphics[scale=0.41, trim=0.5cm 0cm 1.5cm 1.3cm, clip=True
]{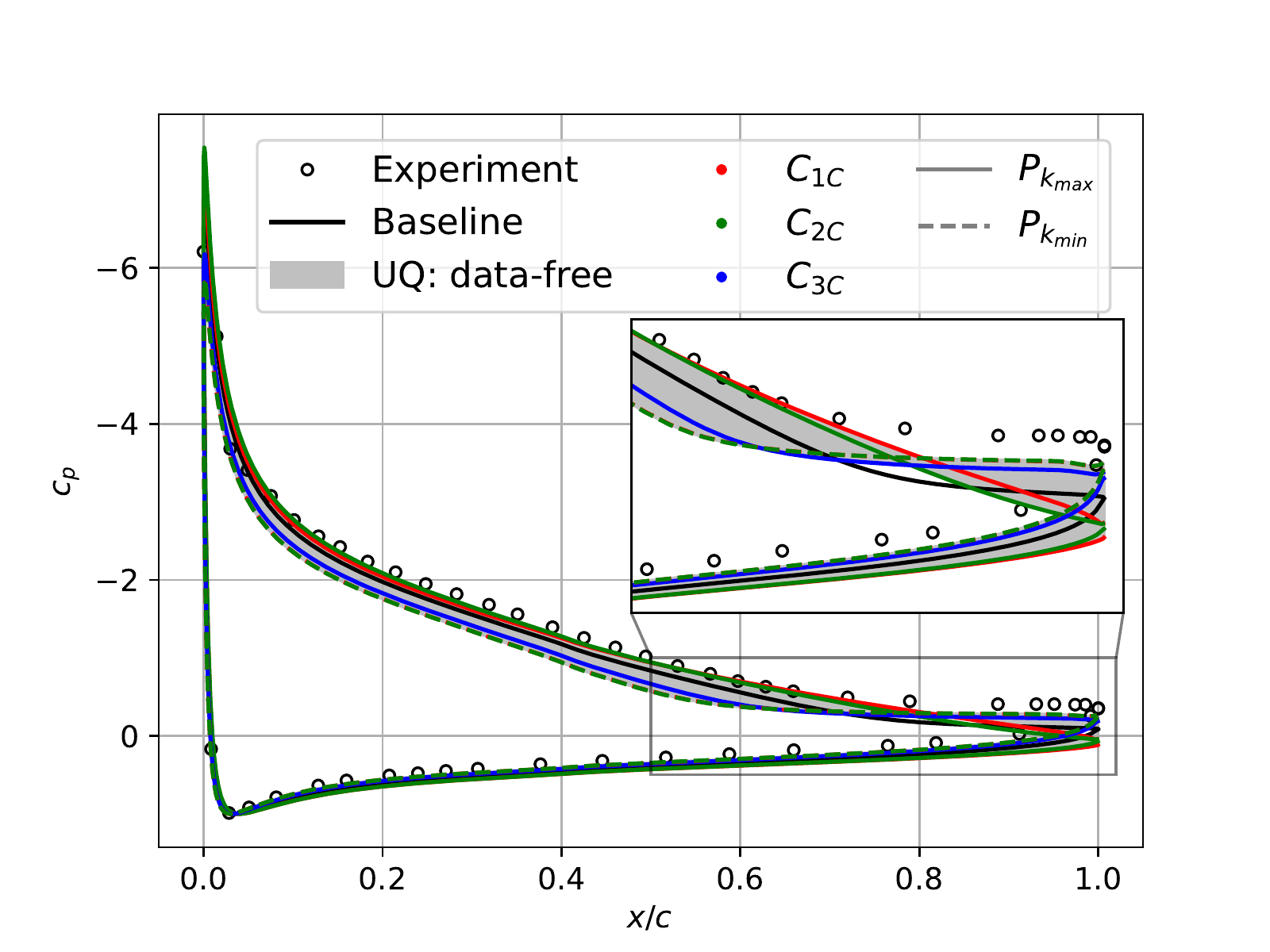}\label{dataFreeCP}}}
\mbox{
\subfigure[Friction coefficient $c_f$]{\includegraphics[scale=0.41, trim=0.4cm 0cm 1.5cm 1.3cm, clip=True]{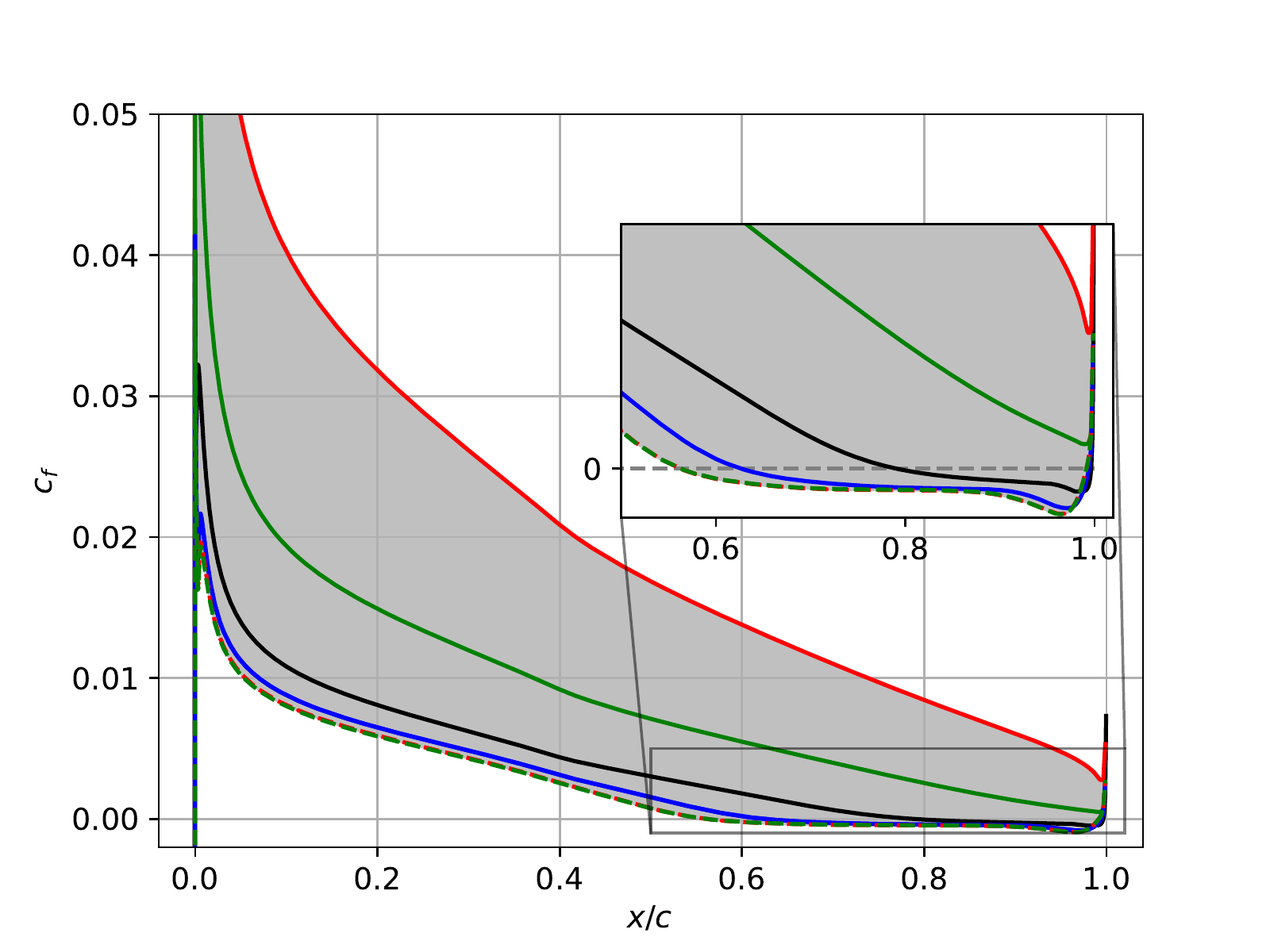}\label{dataFreeCF}}}
\vspace{-0.5\baselineskip}
\caption{Surface quantities for flow around NACA 4412 including data-free evaluation of uncertainty estimates for Menter SST $k-\omega$ turbulence model; legend in (a) applies also for (b)}
\label{Bild:dataFree_naca}
\end{center}
\end{figure}

\noindent The flow around NACA 4412 at $Re_c = 1.52 \cdot 10^6$, $Ma = 0.09$ and an angle of attack of $13.87^{\circ}$ demonstrates the general framework of the UQ perturbation approach presented in \Cref{sec_eigenspaceFramework}.
Before analyzing the actual perturbed solutions and derived UQ estimates, the general performance of the Menter SST $k-\omega$ turbulence model is discussed briefly. 
The baseline simulation is in accordance with the presented RANS solutions using the identical turbulence model provided by NASA's turbulence modeling resource site \cite{NASATurbulenceModelling}.
Similar to NASA's observations, when conducting steady simulation with TRACE on the given grid resolution, $\mathrm{CFL}=1$ has to be used in order to reach a fully converged steady-state solution.
The main difference in comparison with the experimental surface pressure measurements conducted by Coles and Wadcock \cite{Wadcock1979} can be observed at the trailing edge of the suction side. The pressure coefficient, shown for example in \Cref{dataFreeCP}, is defined as
\begin{linenomath}
\begin{equation}
    \label{eq:cp}
    c_p = \frac{p-p_{\mathrm{inf}}}{\frac{1}{2}\rho_{\mathrm{inf}}U_{\mathrm{inf}}^2} \ \text{,}
\end{equation}
\end{linenomath}
while the reference quantities are the ones at infinity $U_{\mathrm{inf}}=31.2 \ \mathrm{m/s}$, $p_{\mathrm{inf}}=76914.1 \ \mathrm{Pa}$ and $\rho_{\mathrm{inf}}=0.9 \ \mathrm{kg/m^3}$ (far-field boundary condition). Although the reference velocity is evaluated at different locations in the experiment, we apply the far-field freestream velocity instead based on best practice guidelines and in order to retain similar CFD results compared to the NASA findings \cite{NASATurbulenceModelling}.\\
The data-free uncertainty estimates, presented in \Cref{Bild:dataFree_naca} are the result of perturbed turbulence model simulations using a relative perturbation magnitude $\Delta_B = 1.0$, since there is no justifiable physical reason to reduce the amount for targeting the extreme states of turbulence \cite{Emory}.
Unfortunately, as already discussed in \Cref{sec:discussionRestrictions}, the perturbed simulations, trying to minimize the turbulent production, come along with stability issues in terms of convergence or even completely diverge from steady-state solutions. An appropriate moderation factor $f$ has to be adjusted to retain acceptable, converged simulations.
Besides examining the overall residuals of each simulation, we evaluate the mean blade force in $y$-direction to distinguish between an unacceptable unstable and an acceptable converged solution. Based on physical experience and consequent intuition, a threshold of 2\%  relative standard deviation with respect to the mean of the overall blade force in $y$-direction is chosen.

\begin{table}[t]
  \caption{Moderation factor $f$ for every perturbed simulation of NACA 4412 UQ estimation. Application of $\Delta_B \leq 1$ in the data-driven approach necessitates a distinction between $P_{k_\textrm{min}}$ and $P_{k_\textrm{max}}$ for 3C.}
  \label{tab:moderationFactorsNaca}
  \hspace*{-1.2cm}
  \centering
  {\footnotesize
  \begin{tabular}{c c c c c c c }
    \toprule
                       &         \multicolumn{5}{c}{Data-free}  &          \\
    \cmidrule(r){2-7}
    Perturbed simulation:          & $(\mathrm{1C}, P_{k_\textrm{max}})$     & $\mathrm{1C}, P_{k_\textrm{min}})$ & $(\mathrm{2C}, P_{k_\textrm{max}})$ & $(\mathrm{2C}, P_{k_\textrm{min}})$& $\mathrm{3C}$&\\
    \midrule
    moderation factor                 & 1.0  & 0.03  & 1.0 & 0.05 & 0.1 & \\
    \midrule
                       & \multicolumn{6}{c}{Data-driven} \\
    \cmidrule(r){2-7}
    Perturbed simulation:        & $(\mathrm{1C}, P_{k_\textrm{max}})$     & $(\mathrm{1C}, P_{k_\textrm{min}})$ & $(\mathrm{2C}, P_{k_\textrm{max}})$ & $(\mathrm{2C}, P_{k_\textrm{min}})$& $(\mathrm{3C}, P_{k_\textrm{max}})$&$(\mathrm{3C}, P_{k_\textrm{min}})$\\
    \midrule
    moderation factor                 & 1.0  & 0.2  & 1.0 & 0.2 & 1.0 &  0.2\\
    \bottomrule
  \end{tabular}
  }
\end{table}

Put simply, as one is increasing the moderation factor $f$ for simulations minimizing the turbulent production term ($P_{k_\textrm{min}}$), the standard deviation of the mean blade force rises. In our investigations, a high-level python script (see \Cref{Bild:TRACEuq}) is aplied to march the moderation factor as high as possible (by steps of 0.1 for $f \in [0.1,1.0)$ and by steps of 0.01 for $f \in [0,0.1)$). As the designated solutions may still contain small variations, we instrument probes on the airfoil surface and averagelav the solution in order to get the mean for QoI. Due to the described convergence issues only a fraction of the perturbed Reynolds stress tensor can be utilized to update the Navier-Stokes equations for the NACA 4412 simulations (see moderation factors in \Cref{tab:moderationFactorsNaca}). To the authors knowledge these low values are in accordance with the implementation of the eigenspace perturbation in the solver suite SU2, as their default value for the moderation factor is 0.1 \cite{SU2Code}. 

\Cref{Bild:dataFree_naca} presents the uncertainty estimates based on the perturbed target states, using the moderation factors presented in \Cref{tab:moderationFactorsNaca}. 
As discussed above, the baseline Menter SST $k-\omega$ simulation shows significant deviation for the prediction of the separation zone close to the trailing edge. The results of $(\mathrm{1C}, P_{k_\textrm{max}})$ and $(\mathrm{2C}, P_{k_\textrm{max}})$ minimize this gap on the suction side for $x/c<0.7$, whereas perturbed simulations minimizing the turbulent production term by modifying the eigenvectors of the Reynolds stress tensor predict an increased static pressure on the suction side for $x/c>0.7$. 
Targeting for the $\mathrm{3C}$ turbulent state with $\Delta_B =1$ results in minimizing the turbulent production term as well \cite{Cook}, which can be seen on the presented surface quantities. Reduced turbulent kinetic energy moves the separation zone towards the front of the airfoil, indicated by the friction coefficient
\begin{linenomath}
\begin{equation}
    \label{eq:cf}
    c_f = \frac{\tau_w}{\frac{1}{2}\rho_{\mathrm{inf}}U_{\mathrm{inf}}^2}
\end{equation}
\end{linenomath}
in \Cref{dataFreeCF}.

In contrast, the boundary layers of the perturbed solutions (1C, $P_{k_\textrm{max}}$) and (2C, $P_{k_\textrm{max}}$) reveal significant increased momentum transfer into the viscous sublayers, inducing complete suppression of the separation bubble.

As already mentioned in \Cref{sec:discussionRestrictions}, the estimated uncertainty bounds by the perturbation framework are only aiming for the extreme state of turbulence in terms of Reynolds stress tensor's shape and orientation, which may not necessarily need to coincide with extreme state of some QoI.
In the range of $0.72 < x/c < 0.82$ the baseline solution lies outside of the determined, grey shaded, UQ estimate for the pressure coefficient.

\begin{figure}[!h]
\begin{center}
\mbox{
\subfigure[Data-driven relative perturbation magnitude evaluated for 1C target state]{\includegraphics[scale=0.28, trim=0.5cm 0cm 0.7cm 0.1cm, clip=False]{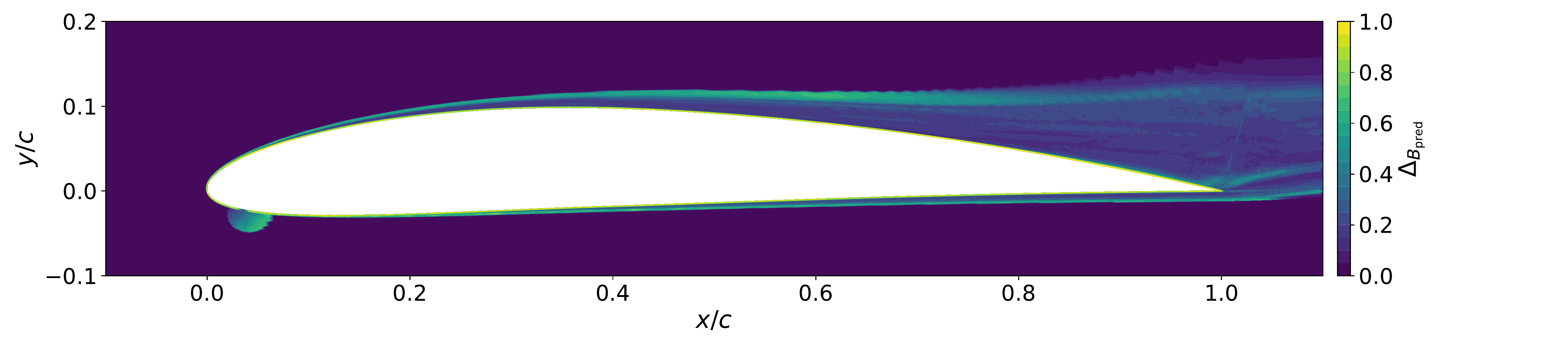}\label{Naca4412C1}}}
\mbox{
\subfigure[Data-driven relative perturbation magnitude evaluated for 2C target state]{\includegraphics[scale=0.28]{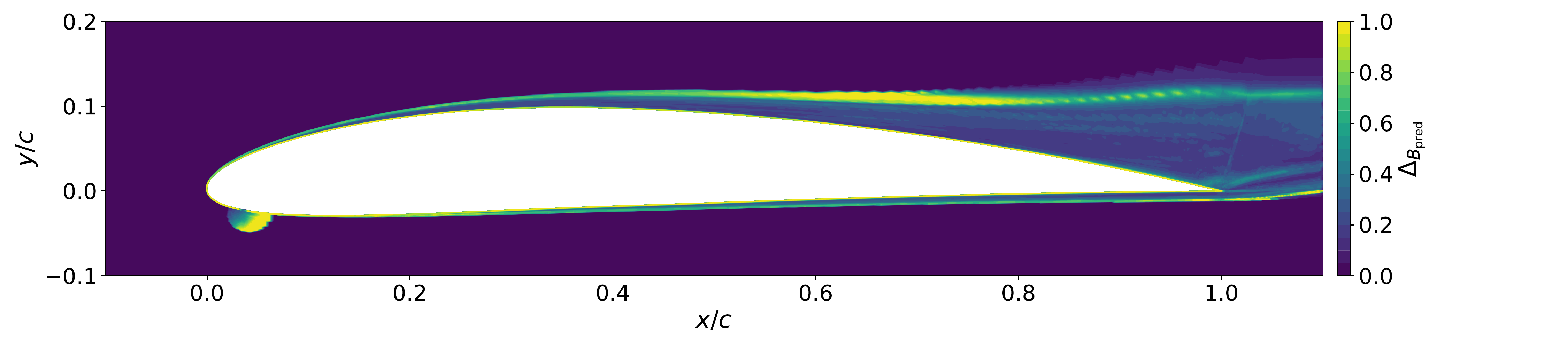}\label{Naca4412C2}}}
\mbox{
\subfigure[Data-driven relative perturbation magnitude evaluated for 3C target state]{\includegraphics[scale=0.28]{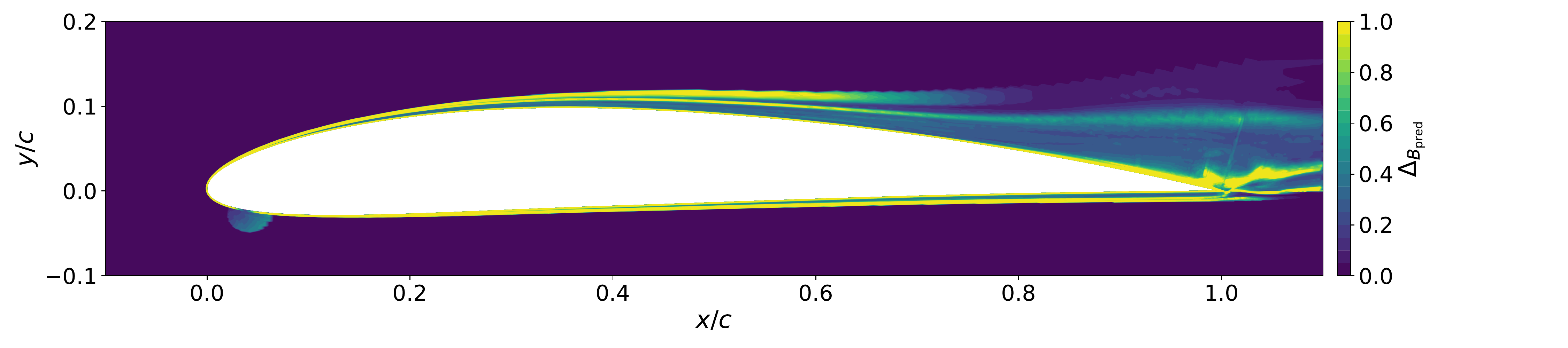}\label{Naca4412C3}}}
\vspace{-0.5\baselineskip}
\caption{Comparison of the effect of identical model predicted perturbation magnitude $p$ on the relative perturbation magnitude $\Delta_B$}
\label{Bild:NACA4412DeltaBEvaluated}
\end{center}
\end{figure}

\begin{figure}[!h]
\begin{center}
\mbox{
\subfigure[Pressure coefficient $c_p$]{\includegraphics[scale=0.41, trim=0.5cm 0cm 1.5cm 1.3cm, clip=True
]{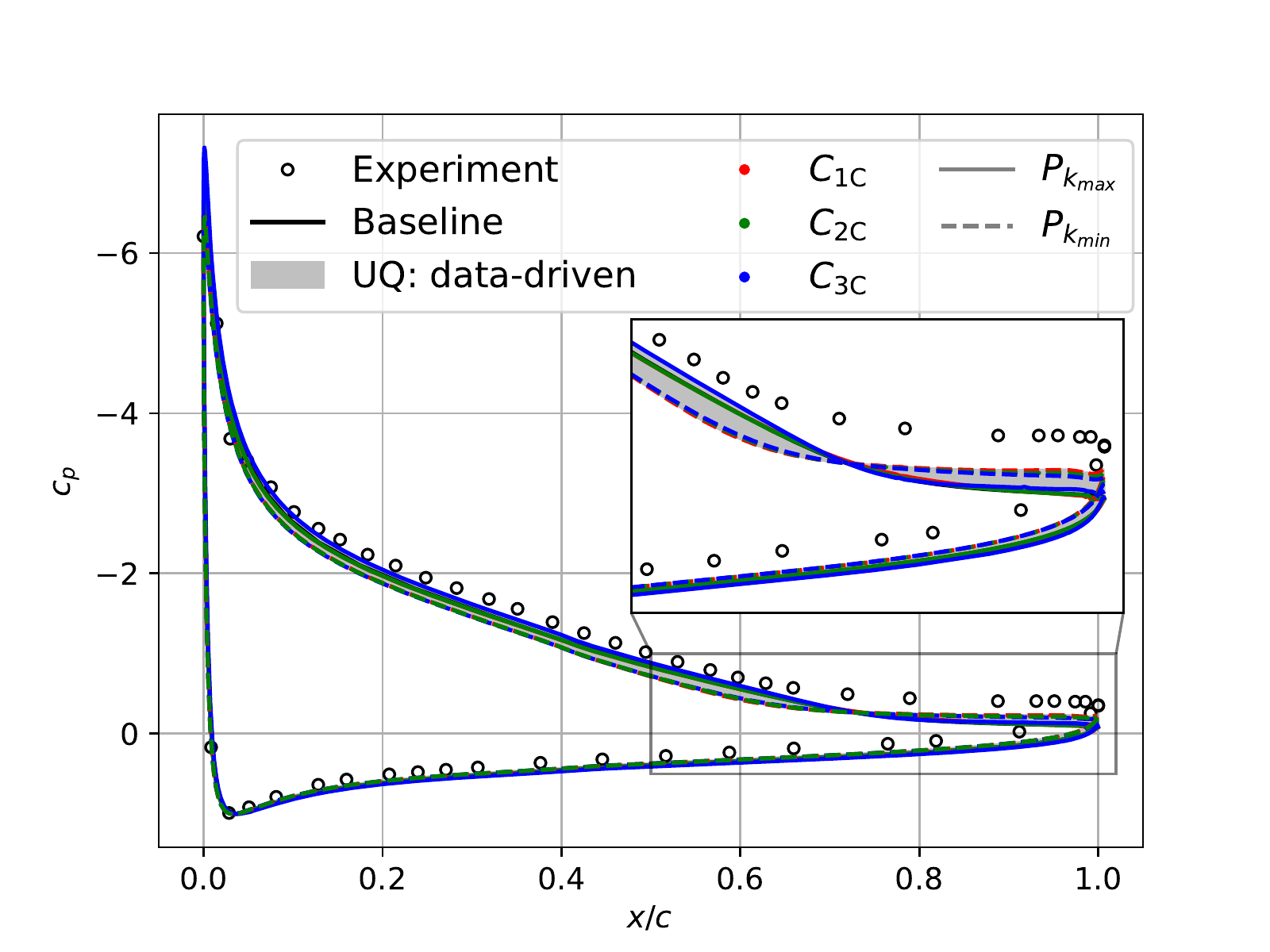}\label{dataDrivenCP}}}
\mbox{
\subfigure[Friction coefficient $c_f$]{\includegraphics[scale=0.41, trim=0.4cm 0cm 1.5cm 1.3cm, clip=True]{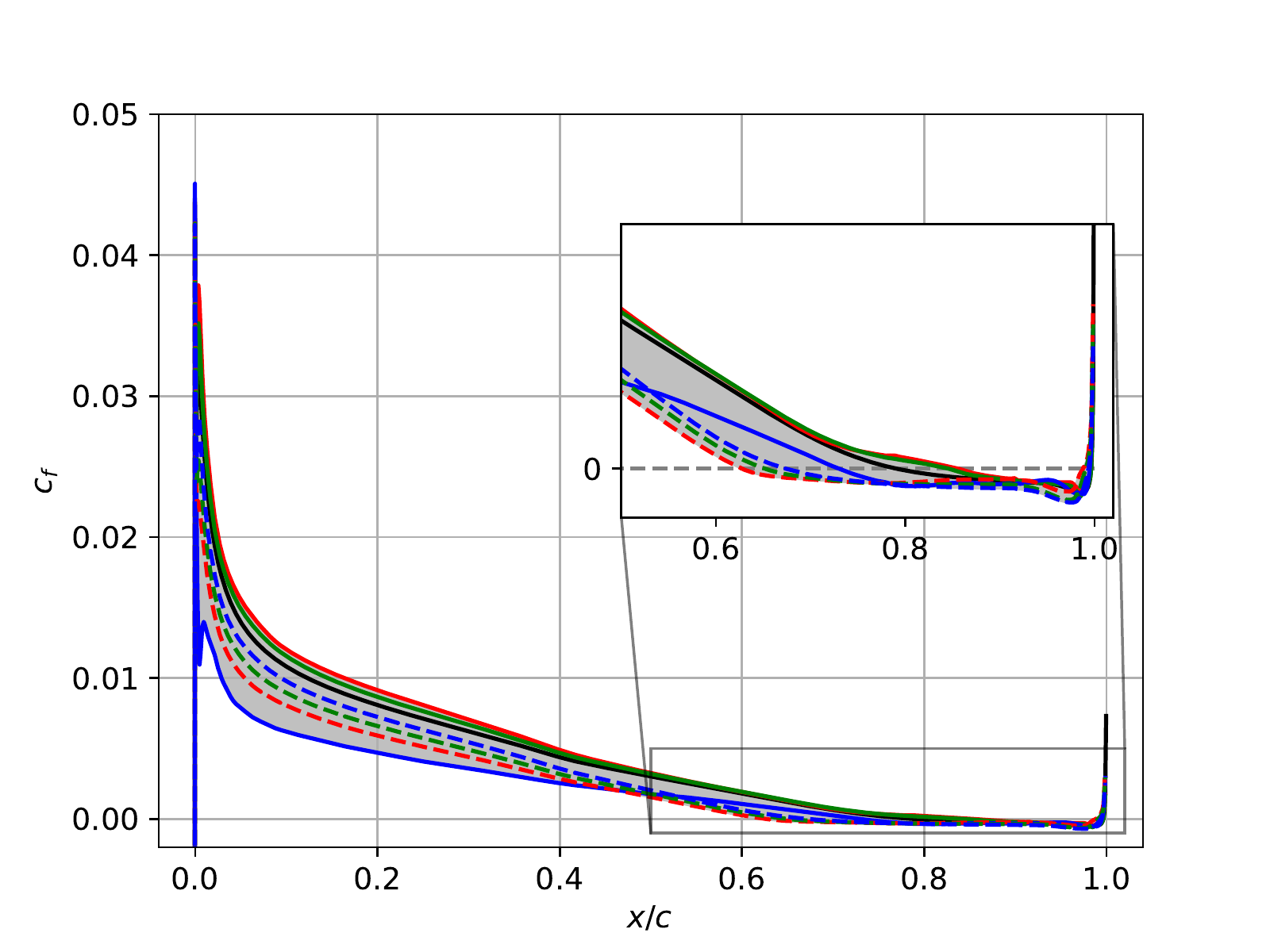}\label{dataDrivenCF}}}
\vspace{-0.5\baselineskip}
\caption{Surface quantities for flow around NACA 4412 including data-driven evaluation of uncertainty estimates for Menter SST $k-\omega$ turbulence model; legend in (a) applies also for (b)}
\label{Bild:dataDriven_naca}
\end{center}
\end{figure}

The random forest predicted perturbation magnitude $p$ (see \Cref{Naca4412P}), is forward propagated towards the same three limiting states as in the data-free approach, as described in \Cref{sec_dataDriven}.
This two-dimensional distribution of $p$ is used to determine the respective $\Delta_B$ for each target state (see \Cref{Bild:NACA4412DeltaBEvaluated} for 1C, 2C and 3C).
Due to the fact, that the unperturbed RANS solution data points are distributed along the plane strain line, the spatially averaged relative distance $\Delta_B$ is highest for the the simulations targeting the isotropic corner (3C), followed by the two-component corner (2C) and the one-component corner (1C). 
In order to reach an acceptable steady-state solution for each perturbed simulation the moderation factor $f$ is adjusted in the same manner as discussed above for the data-free procedure. Since the overall perturbation is weaker than using $\Delta_B=1$, the moderation factor could be increased (see \Cref{tab:moderationFactorsNaca}).\\
The estimated uncertainty bands for the surface quantities, shown in \Cref{Bild:dataDriven_naca}, become narrower. Especially the uncertainty estimates for the pressure coefficient based on $(\mathrm{1C},P_{k_\textrm{max}})$ and  $(\mathrm{2C},P_{k_\textrm{max}})$ are very close to the baseline solution. As the overproduction of turbulent kinetic energy for the data-free 1C and 2C cases disappears, all data-driven perturbed simulations feature a separation zone on the suction surface.

Last but not least, the fact, that none of the presented UQ estimates envelopes the experimental surface pressure measurements, needs to be discussed. To start with, as discussed already in the beginning of this section, the CFD setup seems to come along with certain weaknesses. 
Moreover, the underlying intention of applying the UQ perturbation framework is not to include certain high-fidelity data, whether it originates from experiments or scale-resolving simulation, into its resulting envelope. The methodology seeks to produce limiting states of the Reynolds stress tensor, propagates these states and results in modified QoI. But there is no need that accurate data points for some QoI have to be covered by the uncertainty estimates resulting from the Reynolds stress tensor perturbation framework. 
Therefore, we disagree with some recent publications (e.g. \cite{Eidi}), where the quality of the eigenspace perturbation framework is judged by covering certain QoI. As the uncertainty envelopes do not represent confidence or strict intervals at all \cite{Mishra2020Design}, we encourage to validate the uncertainty estimates with respect to the underlying physical concept of the eigenspace perturbation framework. Analyzing simulation results related to the perturbed state of the Reynolds stress tensor should be chosen over evaluating the coverage of certain reference data.
The main reason is that the perturbation framework is only able to account for structural uncertainties limited to the chosen RANS turbulence model in the CFD-solver. However, other sources of uncertainties related to RANS simulations are not considered by the eigenspace perturbations, such as: 
\begin{itemize}
\item approximation of 3D geometries with 2D CFD setups
\item neglecting of geometry and manufacturing tolerances
\item choice of boundary conditions, which must not necessarily coincide with reference conditions
\item assuming steady-state flow conditions, when flow might be already unsteady in reality.
\end{itemize}
Consequently, it cannot be expected, that the perturbed simulations envelope experimental or reference data.\\
The described machine learning procedure only accounts for the spatially varying deviation in Reynolds stress anisotropy, since the turbulence model's uncertainties are not uniform across the computational domain. The impact of the discrepancy in terms of anisotropy between RANS and scale-resolving data on certain QoI, was not part of the machine learning process. Thus, with reference to the disregarded sources of uncertainty mentioned above, even the data-driven perturbed turbulence model simulations cannot account for adequate entirely enveloping bounds  for selected QoI.

\section{Conclusion and Outlook}
\label{sec_concluison}

\noindent The present work aims to consolidate an arisen method in the field of turbulence model UQ. 
We demonstrate the possibility to estimate uncertainty bounds for turbulence models with state-of-the-art methods in DLR's CFD solver suite TRACE.
The UQ perturbation framework is described extensively, presenting its underlying idea while mentioning its limitations and what it is not able to do for industrial applications. 
Additionally, since TRACE is eminent in turbomachinery industry, designing and implementing a framework to easily conduct uncertainty estimation for turbulence models using TRACE was a major goal of this work.

Moreover, we applied a proposed machine learning strategy to further enhance the interpretability of generated uncertainty estimates by being less conservative and nonetheless physics constrained simultaneously. Our extension of this data-driven eigenvalue perturbation approach is the enlargement of flow cases featuring separated flows, adverse pressure gradient and reattachment for training and testing purposes on the one hand. This enables us to check and verify the appropriate application of trained random forest regression model in-depth. On the other hand, considering eigenvector perturbation of the Reynolds stress tensor on top of the data-driven eigenvalue perturbation was the plausible next step in this specific research field.

In our investigations, we outline tools and methodologies for assessing and analyzing data-driven models, especially in the context of turbulent flows. We address key points in the field of machine learning such as selection of input features, tuning of hyperparameters, judging the model's accuracy in an a posteriori and and an a priori way.

In order to predict the desired target quantity for the selected flow cases by the random forest, we admit, that we might not have to use this abundant number of input features, as described above. This is due to the fact, that the considered cases show certain similarities in terms of input and target quantities. However, if the amount of training data sets increases, covering a wider range of flow phenomena, the machine learning model will likely take advantage of a larger number of input features.

We confirm, that the perturbation approach to account for turbulence model uncertainties, is a purely physics-based, comprehensible framework. Nevertheless, it suffers from reduced convergence or even divergence of steady state solutions. The necessity to moderate certain perturbations by an arbitrary factor, seems unsatisfactory for such an advanced approach. Currently, we are not aware of any other remedy for convergence issues as well, as even the machine learning does not help to overcome this particular issue.
Moreover, we also agree on the underlying idea to account for spatially varying of turbulence model uncertainties by using data-driven methodologies to determine certain areas of high deviations from accurate Reynolds stress anisotropy states. Training a machine learning model to predict an appropriate model-form uncertainty will always help CFD users to get an indication, in which regions the LEVM assumptions might be violated. 
However, we believe, based on our experiences with the data-driven Reynolds stress tensor perturbation framework (which are not only limited to the NACA 4412 flow case), that an enhanced applicability can only be achieved, if the stability issues in terms of convergence for steady state simulation were solved.

\section*{Appendix}
\label{sec:appendix}
\subsection*{Mathematical effect of the moderation factor $f$ in case of pure eigenvalue perturbation}
\label{mathematicalModerationFactor}
\noindent By applying the moderation factor $f \in [0, 1]$ the perturbed Reynolds stress tensor can be expressed as
\begin{linenomath}
\begin{align}
\begin{split}
        \tau_{{ij}_f}^* &= \tau_{ij} + f \left[k \left(a_{ij}^* + \frac{2}{3}\delta_{ij}\right) - \tau_{ij}\right] \\
        &= \left(1-f\right) \tau_{ij} + f \tau_{ij}^* \ \text{,}
\end{split}
\end{align}
\end{linenomath}
where $\tau_{ij}$ is the Reynolds stress tensor, which was calculated based on Boussinesq assumption in step \ref{item:step1} (see \Cref{sec:implementation}).
Based on the perturbed anisotropy tensor, the reconstructed Reynolds stress tensor is indicated by $\tau^*_{ij}$.
The anisotropy tensor related to this perturbed Reynolds stress tensor can be written as
\begin{linenomath}
\begin{align}
\begin{split}
        a_{{ij}_f}^* &= \frac{\tau_{{ij}_f}^*}{k} - \frac{2}{3}\delta_{ij} \\
         &= \frac{\left(1-f\right) \tau_{ij} + f \tau_{ij}^*}{k} + \frac{2}{3}\delta_{ij}  \\
         &= \frac{\left(1-f\right) \left[k\left(a_{ij} + \frac{2}{3}\delta_{ij}\right)\right] + f \left[k\left(a_{ij}^* + \frac{2}{3}\delta_{ij}\right)\right]}{k} - \frac{2}{3}\delta_{ij} \\
         &= \left(1-f\right) \left(a_{ij} + \frac{2}{3}\delta_{ij}\right) + f \left(a_{ij}^* + \frac{2}{3}\delta_{ij}\right) - \frac{2}{3}\delta_{ij} \\
         &= \left(1-f\right) a_{ij} + f a_{ij}^* \ \text{.}
\end{split}
\end{align}
\end{linenomath}

\noindent When just applying eigenvalue perturbation of the anisotropy tensor, $a_{ij}$ and $a_{ij}^*$ share identical eigenvectors. Thus, the eigenvalues of $a_{{ij}_f}^*$ are
\begin{linenomath}
\begin{equation}
        \lambda_{{i}_f}^* = \left(1-f\right) \lambda_{i} + f \lambda_{i}^* \ \text{.}
\end{equation}
\end{linenomath}

\noindent The barycentric weights $C_{{iC}}$, which are used to calculate the barycentric coordinates in Equation \eqref{barycentricMapping}, can be expressed in terms of moderation factor:
\begin{linenomath}
\begin{align}
\begin{split}
\label{eq_c1cf}
        C_{\mathrm{1C}_f}^* &= \frac{1}{2} \left[\lambda_{1_f}^*-\lambda_{2_f}^*\right] \\
        &= \frac{1}{2} \left[\left(1-f\right) \lambda_{1} + f \lambda_{1}^* - \left(1-f\right) \lambda_{2} + f \lambda_{2}^*\right] \\
        &= \left(1-f\right) \frac{1}{2} \left(\lambda_{1} - \lambda_{2}\right)+f\frac{1}{2} \left(\lambda_{1}^*-\lambda_{2}^*\right)\\
        &= \left(1-f\right) C_\mathrm{1C} +f C_\mathrm{1C}^*
\end{split}
\end{align}
\end{linenomath}
\begin{linenomath}
\begin{align}
\begin{split}
\label{eq_c2cf}
        C_{\mathrm{2C}_f}^* &= \lambda_{2_f}^*-\lambda_{3_f}^* \\
        &= \left(1-f\right) \lambda_{2} + f \lambda_{2}^* - \left(1-f\right) \lambda_{3} + f \lambda_{3}^* \\
        &= \left(1-f\right) \left(\lambda_{2} - \lambda_{3}\right)+f\left(\lambda_{2}^*-\lambda_{3}^*\right)\\
        &= \left(1-f\right) C_\mathrm{2C} +f C_\mathrm{2C}^*
\end{split}
\end{align}
\end{linenomath}
\begin{linenomath}
\begin{align}
\begin{split}
\label{eq_c3cf}
        C_{\mathrm{3C}_f}^* &= \frac{3}{2}\lambda_{3_f}^* + 1 \\
        &= \frac{3}{2}\left(1-f\right) \lambda_{3} + \frac{3}{2} f \lambda_{3}^* +(1-f)+f  \\
        &=  \left(1-f\right)\left(\frac{3}{2}\lambda_{3} + 1\right)+f\left(\frac{3}{2}\lambda_{3}^*+1\right)\\
        &= \left(1-f\right) C_\mathrm{3C} +f C_\mathrm{3C}^*
\end{split}
\end{align}
\end{linenomath}

\noindent The perturbed barycentric coordinates $\mathbf{x}_f^*$ (modified by the moderation factor $f$) can be written using Equations \eqref{eq_c1cf}, \eqref{eq_c2cf} and \eqref{eq_c3cf} as
\begin{linenomath}
\begin{align}
\label{eq:xf^2}
\begin{split}
    \mathbf{x}_f^* &= \mathbf{x}_\mathrm{1C} C_{{1C}_f}^*+\mathbf{x}_{\mathrm{2C}_f}^* + \mathbf{x}_{3C} C_{\mathrm{3C}_f}^*\\
        &= \mathbf{x}_\mathrm{1c} \left[\left(1-f\right) C_\mathrm{1C} +f C_\mathrm{1C}^*\right]+\mathbf{x}_\mathrm{2c}\left[\left(1-f\right) C_\mathrm{2C} +f C_\mathrm{2C}^*\right] + \mathbf{x}_{3c} \left[\left(1-f\right) C_\mathrm{3C} +f C_\mathrm{3C}^*\right]\\
        &= \left(1-f\right)  \left(\mathbf{x}_{1c} C_\mathrm{1C}+\mathbf{x}_{2c} C_\mathrm{2C} + \mathbf{x}_{3c} C_\mathrm{3C}\right) + f \left(\mathbf{x}_{1c} C_{\mathrm{1C}}^*+\mathbf{x}_{2c}C_{\mathrm{2C}}^* + \mathbf{x}_{3c} C_{\mathrm{3C}}^*\right)\\
        &= \left(1-f\right) \mathbf{x} +f \mathbf{x}^* \ \text{.}
\end{split}
\end{align}
\end{linenomath}

\noindent Remembering Equation \eqref{perturbationMagnitude} and rearranging leads to 
\begin{linenomath}
\begin{equation}
\label{perturbationMagnitudeRearranged}
		\mathbf{x}^* = \left(1- \Delta_B\right)\mathbf{x} + \Delta_B \mathbf{x}_{(t)} \ \text{.}
\end{equation}
\end{linenomath}

\noindent The analogy of Equation \eqref{eq:xf^2} and \eqref{perturbationMagnitudeRearranged} reveals the similar effect of adjusting $\Delta_B$ or $f$ in case of only perturbing the eigenvalues of the anisotropy tensor.
Thus, one can rewrite the actual intended location inside the barycentric triangle as a relative distance towards the corners 
\begin{linenomath}
\begin{equation}
		\mathbf{x}_f^* = \left(1- \Delta_B f\right)\mathbf{x} + \Delta_B f \mathbf{x}_{(t)} \ \text{.}
\end{equation}
\end{linenomath}

\section*{Acknowledgments}

The project on which this paper is based was funded by the German Federal Ministry for Economic Affairs and Climate Action under the funding code 03EE5041A.
The authors are responsible for the content of this publication.
Additionally, the authors thank the financial funding of the prescribed work by the internal DLR project SuperCOOL.
%An Acknowledgments section, if used, \textbf{immediately precedes} the %References. Sponsorship information and funding data are included here. The %preferred spelling of the word ``acknowledgment'' in American English is without %the ``e'' after the ``g.'' Avoid expressions such as ``One of us (S.B.A.) would %like to thank\ldots'' Instead, write ``F.~A.~Author thanks\ldots'' Sponsor and %financial support acknowledgments are also to be listed in the %``acknowledgments'' section.

\bibliography{mybibfile}

\end{document}